
%
\def\dofig{1}                                                       
%
%

\ifnum\dofig=1
\input epsf   
\fi           

\global\newcount\meqno
\def\eqn#1#2{\xdef#1{(\secsym\the\meqno)}
\global\advance\meqno by1$$#2\eqno#1$$}
%
\global\newcount\refno
\def\ref#1{\xdef#1{[\the\refno]}
\global\advance\refno by1#1}
\global\refno = 1
\vsize=7.5in
\hsize=5in
\magnification=1200
\tolerance 10000
%
%
%
%
\font\sevenrm = cmr7

\hyphenation{non-over-lapping}

\def\bmb{{\bar\mu}_{\beta}}

\def\lb{\lambda_{\beta}}

\def\lbk{\lambda_{\beta,k}}

\def\mb{\mu_{\beta}}
\def\mbk{\mu_{\beta,k}^2}
\def\mbb{\mu_{\beta}^2}

\def\s#1{{\bf#1}}
\def\sp{ \s p}
\def\sx{\s x}

\def\ssp{\sqrt{\sp^2+\mu^2_{\beta}}}

\def\sumn{\sum_{n=-\infty}^{\infty}}

\def\tn{\tilde n}

\def\pmb#1{\setbox0=\hbox{$#1$}%
\kern-.025em\copy0\kern-\wd0
\kern.05em\copy0\kern-\wd0
\kern-.025em\raise.0433em\box0 }
\medskip

\baselineskip=0.1cm
\medskip
\nobreak
\medskip

\baselineskip 12pt plus 1pt minus 1pt
\vskip 2in
\centerline{\bf CONSISTENCY OF BLOCKING TRANSFORMATIONS}
\medskip
\centerline{\bf IN}
\medskip
\centerline{\bf THE FINITE-TEMPERATURE RENORMALIZATION GROUP}
\vskip 24pt
\centerline{Sen-Ben Liao$^{1,2}$\footnote{$^\dagger$}{electronic address:
senben@phy.ccu.edu.tw}
and Michael Strickland $^3$\footnote{$^\ddagger$}{electronic address:
strickland.41@osu.edu}}
\vskip 12pt
\centerline{\it Department of Physics $^1$}
\centerline{\it National Chung-Cheng University}
\centerline{\it Chia-Yi, Taiwan R. O. C.}
\vskip 12pt
\centerline{\it Laboratory of Theoretical Physics $^{2}$}
\centerline{\it Louis Pasteur University}
\centerline{\it 3 rue de l'Universit\'e 67087 Strasbourg, Cedex, France}
\vskip 12pt
\centerline{\it and}
\vskip 12pt
\centerline{\it Department of Physics $^3$}
\centerline{\it Ohio State University }
\centerline{\it Columbus, Ohio\ \ 43210\ \ \ U.S.A.}
\vskip 1.2in
\centerline{Submitted to {\it Nucl. Phys. B}}
\vskip 24 pt
\baselineskip 12pt plus 2pt minus 2pt
\centerline{{\bf ABSTRACT}}
\medskip
\medskip

The finite-temperature renormalization group is 
formulated via the Wilson-Kadanoff blocking transformation.
Momentum modes and the Matsubara frequencies are coupled by 
constraints from a smearing function which plays the role of
an infrared  cutoff regulator. Using the scalar $\lambda\phi^4$
theory as an example,  we consider four general types of  
smearing functions and show that, 
to zeroth-order in the derivative expansion, they
yield qualitatively the same 
temperature dependence of the running constants and the same
critical exponents within numerical accuracy.

\vskip 24pt
\vfill
\noindent CCU-TH-98-01 \hfill March 1998
\eject

\vfill
\eject

\centerline{\bf I. INTRODUCTION}
\medskip
\nobreak
\xdef\secsym{1.}\global\meqno = 1
\medskip
\nobreak

Finite-temperature renormalization group (RG) techniques provide a powerful 
means of probing the physical behavior of a quantum system interacting with 
an external heat bath.  The presence of the additional length scale 
$\beta=T^{-1}$, the inverse of the temperature, makes the theory more 
complicated since there are now two types of fluctuations, quantum and 
thermal, which compete with one another. In order to apply the RG framework 
to explore issues such as the QCD deconfinement phase transition and the 
evolution of the early universe, one must first understand the subtle interplay 
between these fluctuations. 

To take into account both thermal and quantum fluctuations,
various finite-temperature RG prescriptions have been proposed \ref\lp\
,\ref\others. These RG schemes also
provide an improvement beyond perturbative approximations,
which are known to be plagued by infrared (IR) singularities in the
high-temperature limit as well as in the vicinity of 
criticality. In the present work, we revisit the studies
of Ref. \lp\ which
were based on the Wilson-Kadanoff blocking transformation \ref\wilson\
performed on a manifold
$S^1\times R^d$ in the imaginary-time formalism.  
By introducing a regulating smearing function
$\rho^{(d)}_{\tn, k}({\bf x},\tau_x)$, which simulates the coarse graining in
momentum space, the effective degrees of freedom of the theory are now
characterized by the blocked fields, $\phi_{\tn,k}$,
\eqn\bloc{\phi_{\tn,k}({\bf x},\tau_x)\equiv\int_0^{\beta}d\tau_y\int_{\bf y}
\rho^{(d)}_{\tn,k}({\bf x}-{\bf y}, \tau_x-\tau_y)\phi(\s y,\tau_y)\,,}
where $\phi(x)$ represents the original field variables. While $\tn$
is a discrete cutoff for the Matsubara frequencies $\omega_n=2\pi n/{\beta}$, 
the scale $k$
may be regarded as a generalized IR cutoff for a combination of both
$|\sp|$ and $\omega_n$. 
Varying $k$ infinitesimally then allows us to 
generate a nonlinear differential RG equation for the blocked action
${\widetilde S}_{\beta,k}[\Phi]$, which is defined by
\eqn\cactt{ e^{-\widetilde S_{\beta,k}[\Phi(\sx)]}=\int_{\rm periodic} 
D[\phi]\prod_{\sx}
\delta(\phi_k(\sx)-\Phi(\sx))e^{-S[\phi]}\,.}
The blocked action provides a smooth interpolation 
between the bare action $S_{\Lambda}[\Phi]$ 
and the quantum effective action which generates the 
one-particle-irreducible graphs as $k$ is lowered from the ultraviolet (UV)
cutoff, $\Lambda$, to zero.
We have taken the field average of a given block $\Phi(\sx)$ to coincide with
the slowly varying background. 
Since the flow equation for ${\widetilde S}_{\beta,k}[\Phi]$ is too
complicated to be solved exactly, we make an expansion in 
powers of derivatives:
\eqn\derivs{{\widetilde S}_{\beta,k}[\Phi]=\int_0^{\beta}d\tau\int_{\s x}
\Biggl\{
{Z_{\beta,k}(\Phi)\over 2}(\partial_{\tau}\Phi)^2+{Z_{\beta,k}(\Phi)\over 2}
(\nabla\Phi)^2+U_{\beta,k}(\Phi)+O(\partial^4)\Biggr\}\,.}
The theory is then parameterized by a set of coupled nonlinear partial
differential RG equations for the wavefunction renormalization constant,
$Z_{\beta,k}(\Phi)$, the blocked potential, $U_{\beta,k}(\Phi)$ and
higher-order contributions. The behavior of the theory
at any arbitrary $k$ and $T$ can then be studied by 
analytically or numerically solving these coupled equations.
We shall designate
this type of RG prescription as the momentum RG, or MRG. 
 
Notice that the functional form of
$\rho^{(d)}_{\tn,k}({\bf x}, \tau_x)$ in MRG is not unique;
different choices 
will change the way in which the irrelevant,
short-distance degrees of freedom are
eliminated in course of blocking. In the finite-temperature setting,
how the two types of modes, $\omega_n$ in $S^1$ and 
$|\sp|$ in $R^d$, are block averaged is controlled 
by $\rho^{(d)}_{\tn,k}({\bf x}, \tau_x)$. 
In Fourier space, $\rho^{(d)}_{\tn,k}({\bf p}, \omega_n)$ modifies the
bare propagator as
\eqn\propag{{1\over{\sp^2+\omega_n^2+\mu^2_R}}\longrightarrow
{{\rho^{(d)}_{\tn,k}({\bf p}, \omega_n)}\over{\sp^2+\omega_n^2+\mu^2_R}}\,.}
Thus, one may visualize the following possibilities: (i) Blocking is 
performed in only one of the two subspaces, i.e., 
with $\rho_{\tilde n}(\tau_x)$, or 
$\rho_k^{(d)}(\sx)$; (ii) both $\omega_n$ and $|\sp|$ are 
blocked, but with two smearing functions
$\rho_{\tilde n}(\tau_x)$ and $\rho_k^{(d)}(\sx)$ that are independent
of one another; and (iii) a constraint exists in the
smearing function for $\omega_n$ and $|\sp|$ which allows them
to be coupled and eliminated in some coherent manner. 
The trivial case of no blocking simply implies the absence of any RG 
transformation. Scenario (iii) is the most interesting case because it 
allows us to examine the interplay between the quantum and thermal 
fluctuations in the two spaces. In this paper smearing functions 
corresponding to the above categories are constructed explicitly.
In particular, we study, in Fourier space, 
$\rho_k^{(d)}(\sp)=\Theta(k-|\sp|)$ for category (i), 
$\rho_{\tilde n,k}^{(d)}(\sp,\omega_n)=\Theta(k-|\sp|)\Theta(\tilde n-|n|)$
for (ii), and 
$\rho_k^{(d)}(\sp, \omega_n)=\Theta(k-\sqrt{\omega_n^2+|\sp|^2})$
and $\rho_k^{(d)}(\sp, \omega_n)=\Theta(k-|\omega_n|-|\sp|)$ for (iii). 
These smearing functions differ only in their treatment of the
coupling between $|\sp|$ and the $n\ne 0$ sector. Thus,
by comparing the numerical solutions of the corresponding RG equations,
one can gain insight into the importance of the feedback 
between $\omega_n$ and $|\sp|$.  

To derive the MRG equations, we lower $k$ infinitesimally from
$k$ to $k-\Delta k$, where $\Delta k$ is the characteristic thickness of
the RG shell that contains the modes to be eliminated. The
ratio $\kappa=\Delta k/k$ represents the fraction of the modes
that are eliminated in each blocking step. When the smearing function
involves $n$ explicitly, the resulting RG becomes piece-wise continuous
due to the discreteness in the Matsubara frequencies.
However, we shall find that in all the cases considered, qualitatively the
same results are obtained for the temperature dependence of the running
parameters.  The quantitative difference 
stems from the boundary conditions on the
RG flow when the Matsubara modes are blocked. 

We also show that all four methods yield the same critical
exponents to within 0.5\% when $U_{\beta,k}(\Phi)$
is expanded in a power series in $\Phi$ near criticality. 
Thus, it may be concluded that
the feedbacks between $|\sp|$ and the $n\ne 0$ sector can be treated 
perturbatively in the high $T$ regime. Although 
scaling patterns in the
IR regime differ in all four cases, direct correspondence may 
nevertheless be established and equivalence may be
recovered in some cases.

We emphasize that $\kappa$ must always remain small in order to trace out 
the true RG evolution. A large $\kappa$ causes poor 
tracking of the RG trajectory.
To illustrate this point,
we examine the high-temperature behavior of the  
quartic coupling constant of the scalar
$\lambda\phi^4$ theory in the symmetric phase. 
It has been shown that $\lambda_{\beta}=\lambda_{\beta,k=0}$ decreases with 
increasing $T$ but approaches a constant 
using $\rho_k^{(d)}(\sp)=\Theta(k-|\sp|)$ \ref\ms. However, if one employs the 
``Matsubara shell assumption'' with $\Delta k=2\pi/{\beta}$ , i.e.,
the thickness of the RG shell is chosen to be equal to
the spacing between two adjacent Matsubara frequencies, an incorrect
high-temperature behavior will be obtained. 
The result is due to the fact that as $\Delta k$ becomes increasing large 
in the high $T$ limit, 
$\kappa$ is no longer small and the flow
then deviates significantly from its true trajectory.

The organization of the paper is as follows: In Sec. II we give four
examples of the MRG smearing functions and describe the details of
how the corresponding RG evolution equations are generated. We 
compare and contrast the manner in
which $\omega_n$ and $|\sp|$ are coupled.
Numerical solutions for the MRG equations are presented
in Sec. III. We demonstrate once more how
$\lambda_{\beta,k=0}$ decreases 
with increasing $T$ and may be approximated by a constant in the 
infinite-temperature limit.  
Results extracted from the Matsubara shell assumption are also included to
illustrate its failure in the high-temperature limit.
Sec. IV is
reserved for summary and discussions. In Appendix A we present a 
complementary RG scheme, the temperature RG (TRG), which is formulated
using $T$ as the running parameter. Its consistency with MRG is
demonstrated. In Appendix B the connection between the 
the finite-temperature behavior of the running parameters and the 
fixed points is established. In particular, 
we identify all
the fixed points accessible by both the MRG and the TRG schemes and show that 
the existence of an infinite-temperature Gaussian fixed point
rules out the possibility of $\lambda_{\beta,k=0}$ increasing with $T$.

\medskip
\medskip
\centerline{\bf II. MRG SMEARING FUNCTIONS}
\medskip
\nobreak
\xdef\secsym{2.}\global\meqno = 1
\medskip
\nobreak

As described in the Introduction, under blocking transformations
an IR cutoff $k$ is introduced 
into the theory by defining the coarse-grained fields \bloc\
via a
smearing function $\rho^{(d)}_{\tn,k}(\sx, \tau_x)$. 
Here $\tn$ differentiates between the high and the low Matsubara 
frequencies, and $k$ provides a
separation between the slowly varying background fields and the
fast-fluctuating modes which are to be integrated over. Below we consider
four smearing functions, each of which has a different 
coupling between $\omega_n$ and $|\sp|$.

\medskip
\noindent{(1)} $\rho_k^{(d)}({\s p})=\Theta(k-|{\s p}|)$:
\medskip

The simplest blocking is to choose a sharp cutoff in the $R^d$ submanifold,
leaving $S^1$ unconstrained.
In position space, this smearing function has the form
\eqn\smear{\rho_k^{(d)}({\bf x})=\int_{|{\bf p}| < k}
e^{i{\bf p}\cdot{\bf x}},}
which approaches $\delta^d(\sx)$ as $k\to\infty$. This
limit corresponds to a completely ``unblocked'' bare system characterized
by the original unrenormalized fields. The blocked fields take on the 
form \ms\
\eqn\blok{\phi_k({\bf x},\tau_x)=\int_{\bf y}
\rho^{(d)}_k({\bf x}-{\bf y})\phi(\s y,\tau_x)
={1\over\beta}\sum_{n=-\infty}^{\infty}
\int_{|{\bf p}| < k}e^{-i(\omega_n\tau_x-{\bf p}\cdot{\bf x})}
\phi_n({\bf p}).}
We see that the effect of this smearing is to provide an integration over
$|{\bf p}| > k$ for all $n$. However, with
this choice of $\rho^{(d)}_k(\sx)$, no averaging is done in the $\tau_x$
direction. 

Let us first consider the simple, independent-mode 
approximation where
all the modes greater than $k$ are eliminated at once with no feedback 
being taken into account. The perturbative
one-loop correction to the blocked potential 
reads:
\eqn\fubk{\eqalign{\tilde U^{(1)}_{\beta, k}(\Phi) &={1\over2\beta}\sumn
\int_{\s p}^{'}{\rm ln}\Bigl[\omega_n^2+\sp^2+V''(\Phi)\Bigr]={1\over\beta}
\int_{\s p}^{'}{\rm ln~sinh}\Bigl({\beta\sqrt{\sp^2
+V''(\Phi)}\over 2}\Bigr)\cr
&
={1\over 2\beta}\int_{\s p}^{'}\Biggl\{\beta\sqrt{{\s p}^2+V''(\Phi)}
+2~{\rm ln}\Bigl[1-e^{-\beta\sqrt{\sp^2+V''(\Phi)}}\Bigr]\Biggr\}+\cdots\,,}}
where the prime notation in the integration implies $k < |\sp| < \Lambda$. 
The wavefunction renormalization constant has been set to unity
for simplicity. A direct differentiation of the above expression with 
respect to $k$ then gives
\eqn\ima{k{{d {\tilde U}}_{\beta, k}(\Phi)\over {d k}}=
-{S_dk^d\over 2\beta}\Biggl\{\beta\sqrt{k^2+V''(\Phi)}
+2~{\rm ln}\Bigl[1-e^{-\beta\sqrt{k^2+V''(\Phi)}}\Bigr]
\Biggr\}\,,}
where $S_d=2/{(4\pi)^{d/2}\Gamma(d/2)}$. Unfortunately the equation
has a serious drawback in that it
only accounts for the one-loop contribution and neglects the important
feedback from the higher to the lower modes whose energy scales, 
in principle,
may differ by several orders of magnitude. Since
the effective degrees of
freedom of the theory are not well tracked, it will fail to capture the
relevant physics in the high temperature limit as well as in the
neighborhood of the phase transitions.

The aim of MRG is precisely to take into consideration the important 
higher loop corrections and the couplings between the modes.
In Figure 1 we depict how RG is carried out as an improvement beyond the
independent-mode approximation. At $k=\Lambda$, the system is unblocked and 
no modes are eliminated. As the system undergoes blocking with the IR cutoff
being lowered from $k$ to $k-\Delta k$, the modes contained 
inside the RG shell of thickness $\Delta k$ are integrated out. The
elimination of these fast-fluctuating modes results in a more 
complicated blocked 
potential $U_{\beta,k-\Delta k}(\Phi)$ which contains more 
operators than the previous one, $U_{\beta,k}(\Phi)$. 
The evolution is complete when
$k$ finally reaches zero. By lowering $k$
infinitesimally, the RG equation reads: 
\eqn\ublok{ k{{d U_{\beta,k}(\Phi)}\over{d k}}=
-{S_dk^d\over 2\beta}\sum_{n=-\infty}^{\infty}{\rm ln}\Bigl[\omega_n^2
+k^2+U''_{\beta,k}(\Phi)\Bigr]\,,}
which, upon summing
over $n$, yields
\eqn\rgft{k{{d U}_{\beta, k}(\Phi)\over {d k}}=
-{S_dk^d\over 2\beta}\Biggl\{\beta\sqrt{k^2+U''_{\beta,k}(\Phi)}
+2~{\rm ln}\Bigl[1-e^{-\beta\sqrt{k^2+U''_{\beta,k}(\Phi)}}\Bigr]
\Biggr\}\,.}
The functional forms of Eqs. \ima\ and \rgft\ are remarkably similar,
except for the substitution on the right-hand-side of the latter by the
scale-dependent counterpart, i.e., $V''(\Phi)\to U''_{\beta,k}(\Phi)$.
Though this ``dressing'' seems straight-forward, it has been demonstrated
to have far-reaching
consequences, particularly in the study of
phase transitions and critical phenomena. 
By solving the equations numerically subject to the initial condition
$U_{\beta=\infty,\Lambda}(\Phi) =V_{\Lambda}(\Phi)$, where $V_{\Lambda}(\Phi)$
is the bare potential defined at the UV cutoff $\Lambda$, 
the severe IR divergences encountered in the finite-loop calculations are 
completely removed.  In addition, 
critical exponents can be measured to a remarkable precision through numerical
solutions of the exact renormalization group flow equations \ref\mike.

\ifnum\dofig=1                                                       
\medskip                                                             
                                           
\centerline{\epsfbox{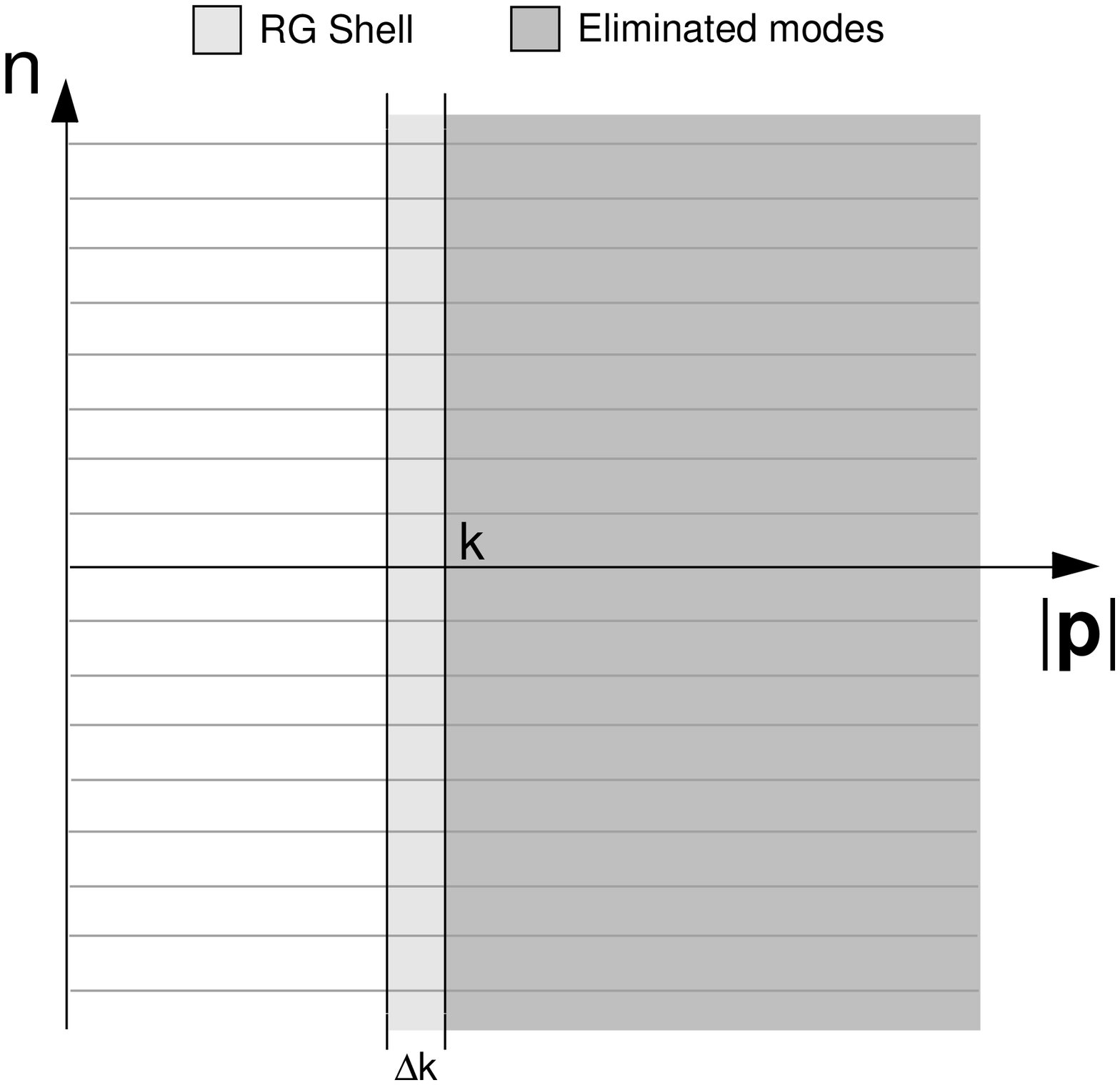}}                                      
\medskip                                                             
{\narrower                                                           
{\sevenrm                                                            
{\baselineskip=8pt                                                   
\centerline{
\itemitem{Figure 1.}                                                 
Schematic diagram of blocking Method 1.}
\bigskip                                                             
}}}                                                                  
\fi                                                                  

It is useful to compare Eq. \rgft\ with the 
zero-temperature result for $D$ dimensions, obtained using a sharp cutoff for
$D$-dimensional momentum integration \ref\sb:
\eqn\ulfe{ k{{d U_k(\Phi)}\over{d k}}=
-{S_Dk^D\over 2}{\rm ln}\Bigl[k^2+U''_k(\Phi)\Bigr].}
In the regime where $\bar\beta\to 0$, 
the contribution from the first term in \rgft\ can be 
neglected and one has
\eqn\ulrg{ k{{d U_{\beta,k}(\Phi)}\over{d k}}=
-{S_dk^d\over 2\beta}{\rm ln}\Bigl[k^2+U''_{\beta,k}(\Phi)\Bigr],}
which is simply the contribution from the $n=0$ mode, as can be seen from
\ublok. Thus, we see clearly that 
the $n=0$ sector alone is sufficient to account for
the $d$-dimensional characteristics.
Upon rescaling of $\Phi\to {\beta}^{-1/2}\Phi$ and $U_{\beta,k}\to
\beta^{-1}U_k$, Eq. \ulrg\ coincides with Eq. \ulfe\ for $D=d$.

On the other hand, in the low $T$ regime where $\bar\beta\to\infty$ (holding
$k$ fixed), a complete summation over $n$
is necessary for recovering the proper zero-temperature
$(d+1)$-dimensional feature of the theory.
The expected RG equation is that of Eq. \ulfe\ with
$D=d+1$. However, in this limit Eq. \rgft\ gives
\eqn\rgff{k{{d U}_k(\Phi)\over {d k}}=
-{S_dk^d\over 2}\sqrt{k^2+U''_k(\Phi)}}
instead. 
A careful comparison of the RG flow governed by Eqs. \ulfe\ and \rgff\ 
shows that they lead to the same renormalized values for the 
vertex functions at $k=0$, but differ in the scalings patterns 
for $k \ll \mu_R$, where $\mu_R$ is the renormalized mass of the theory \ms.
How can we improve our RG to ensure that the desirable IR scaling limits 
below the mass gap are attained? To do so, one must carefully
take into account the couplings between $\omega_n$ and $|\sp|$.  From Figure 1,
it is clear that Method 1 results in a complete summation over
all Matsubara modes within a given RG shell and ignores the couplings between
modes having the same $|\sp|$ but with different $n's$. 
How do we take into consideration this coupling, and how 
important is this to physical applications? 
To answer these questions, we must examine other smearing functions.

\medskip
\medskip

\item{(2)} $\rho_{{\tilde n},k}^{(d)}({\s p},\omega_n)=\Theta(k-|{\s p}|)
~\Theta({\tilde n-|n|})$:
\medskip

As an improvement to account for the systematic feedback between certain
$d$-dimensional momentum modes $|\sp|$ 
and the Matsubara frequencies $\omega_n$, we choose
the smearing function to be $\rho_{{\tilde n},k}^{(d)}
({\s p},\omega_n)=\Theta(k-|{\s p}|)~\Theta({\tilde n-|n|})$, or
\eqn\smea{ \rho_{\tilde n, k}^{(d)}(\sx,\tau_x)=\rho_{\tilde n}(\tau_x)
\rho_k^{(d)}(\s x)
={1\over\beta}\sum_{|n| < {\tilde n}}\int_{|\s p| < k}
e^{-i(\omega_n\tau_x-\s p\cdot\s x)}\,,}
which implies:
\eqn\blc{ \phi_k(\sx,\tau_x)={1\over\beta}
\sum_{|n| < \tilde n}\int_{|\s p| < k}
e^{-i(\omega_n\tau_x-\s p\cdot\s x)}\phi_n(\s p)\,.}
In other words, this smearing 
function provides a summation over $n > |\tilde n
|$ in $S^1$ and an integration over all $|{\sp}| > k$ in $R^d$. From
\eqn\orthg{\delta(\tau_x)={1\over\beta}\sum_{n=-\infty}^{\infty}
e^{-i\omega_n\tau_x}\,,}
one can readily verify again that the limits $k\to\Lambda$ and $\tilde n\to 
\infty$ correspond to $\rho_{\tilde n, k}^{(d)}(\sx,\tau_x)\to
\delta^d(\sx)\delta(\tau_x)$ and the theory coincides with the original
``unblocked'' system. With such a step-by-step elimination,
the couplings between different Matsubara frequencies are also taken into
account \ref\patkos.

It turns out that the
bounds on $n$ can actually be chosen in a more physically motivated
manner. To do so, we notice that just as the high modes with $|\s p| > k$
in $R^d$ are being integrated over, we should also sum over in $S^1$ all
the high Matsubara frequencies with $|\omega_n| > k$, or $|n| >
[\bar\beta/{2\pi}]$, where $\bar\beta=\beta k$ and $[j]$ is the greatest
integer less or equal to $j$. A similar upper bound, $\beta\Lambda/{2\pi}$
on $\omega_n$, or equivalently, $n_{\rm max}=[\beta\Lambda/{2\pi}]$ may
also be imposed. The blocking procedure of this scheme is depicted in
Figure 2. 

\ifnum\dofig=1                                                       
\medskip                                                             
                                           
\centerline{\epsfbox{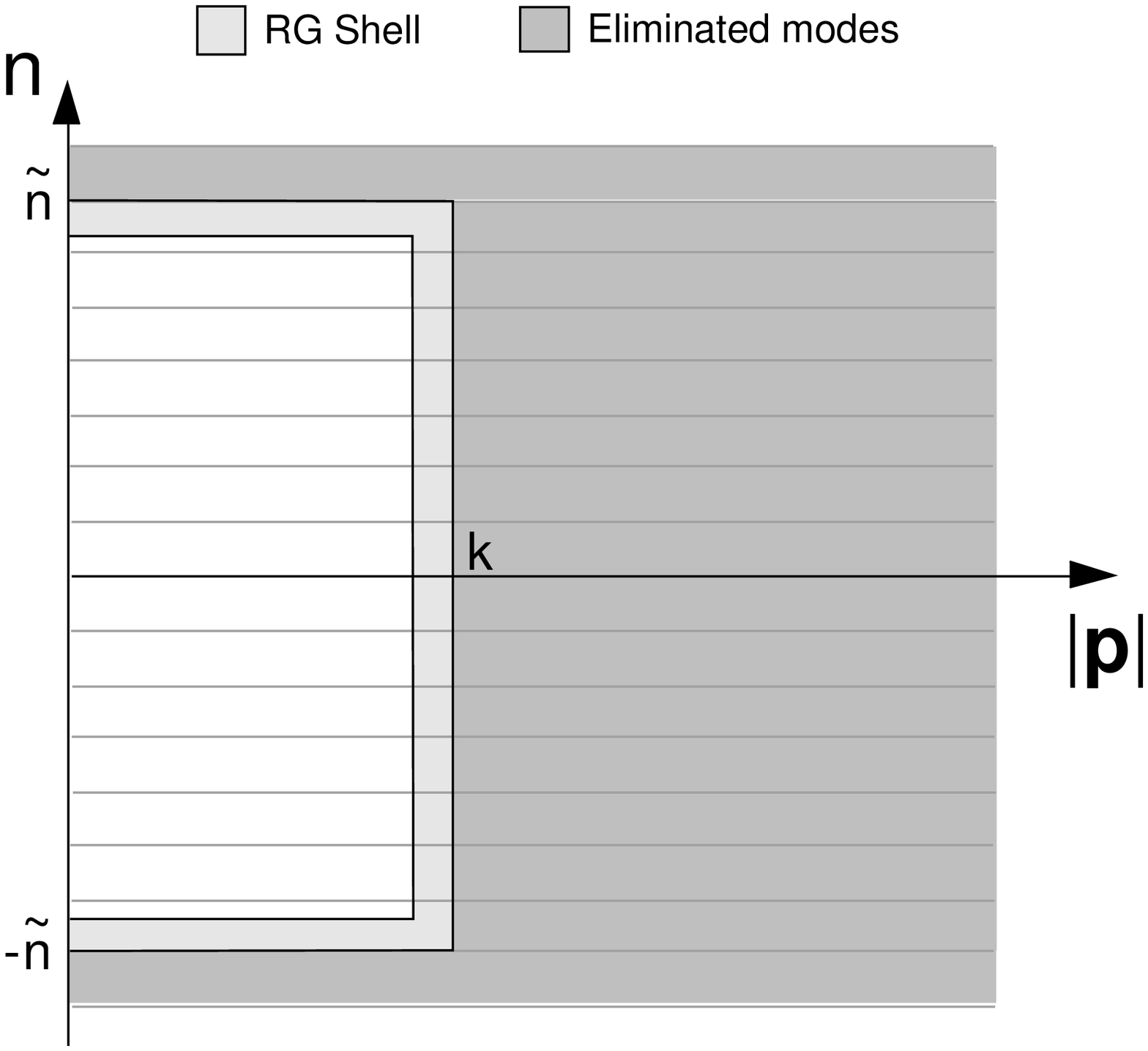}}                                      
\medskip                                                             
{\narrower                                                           
{\sevenrm                                                            
{\baselineskip=8pt                                                   
\itemitem{Figure 2.}
Schematic diagram of blocking Method 2. The thickness of the RG shell
in the $\scriptstyle n$-axis is zero for $\scriptstyle \Delta k \to 0$ 
(exaggerated here), but a Matsubara mode is eliminated after $\scriptstyle 
\cal N$ continuous iterations where 
$\scriptstyle [{\cal N}\beta\Delta k/{2\pi}]=1$.
\bigskip                                                             
}}}                                                                  
\fi         

From the figure, we see that the RG shell on the $\omega_n-|\bf p|$ plane
can be subdivided into three parts: one vertical strip having 
a thickness $\Delta k$
containing modes with $|n|\le [\bar\beta/{2\pi}]$, and two horizontal strips  
with $|\s p| < k$. In the limit
$\Delta k\to 0$, the contribution from the two horizontal strips vanishes, 
as long as $\Delta k$ does not span a Matsubara mode.
It is crucial not to make the Matsubara shell assumption with
$\Delta k=2\pi/{\beta}$ such that all three volumes contribute.
Such an assumption will lead to serious difficulties, particularly in the
high temperature regime where $\Delta k$ becomes so large that the
treatment would be reduced to that of the independent-mode approximation.
Thus, one always makes $\Delta k$ small to ensure that
the RG shell contains as few modes as possible. 
The subtraction of a Matsubara mode takes place only 
after every $\cal N$ continuous iterations of RG in the $|\sp|$
space when ${\cal N}\Delta k=2\pi/{\beta}$ and the condition
$[{\cal N} \beta\Delta k/{2\pi}]=1$ is reached. Similar blocking 
iterations are then repeated until we arrive at $n=0$. 

The above prescription leads to 
\eqn\mtwo{\eqalign{U_{\beta,k-\Delta k}-U_{\beta,k} &=
{1\over 2\beta}\sum_{|n|\le [{\bar\beta\over{2\pi}}]} 
\int_{\sp}\Theta\bigl(|\sp|
-k+\Delta k\bigr)\Theta\bigl(k-|\sp|\bigr)~{\rm ln}\Bigl[\omega_n^2+p^2
+U''_{\beta,k}(\Phi)\Bigr] \cr
&
={S_d\over 2\beta}\sum_{n=-[{\bar\beta\over{2\pi}}]}^{[{\bar\beta\over{2\pi}}]}
\int_{k-\Delta k}^k dp~p^{d-1}~{\rm ln}\Bigl[\omega_n^2+p^2
+U''_{\beta,k}(\Phi)\Bigr],}}
or, in the limit $\Delta k\to 0$,
\eqn\urg{ k{{d U_{\beta,k}(\Phi)}\over{d k}}=
-{S_dk^d\over 2\beta}\sum_{n=-[\bar\beta /{2\pi}]}^{[\bar\beta/{2\pi}]}
{\rm ln}\Bigl[\omega_n^2+k^2+U''_{\beta,k}(\Phi)\Bigr].}
We see that contrary to Eq. \rgft\ in Method 1, the summation over $n$ 
is now restricted to that contained in the shell, namely $|n| \le 
[\bar\beta/{2\pi}]$. This RG coincides with Eq. \ulrg\ as $\bar\beta\to 0$.
On the other hand, when $\bar\beta\to\infty$, one recovers
\ublok\ or \rgft\ which contains both zero and finite-temperature 
sectors for $d+1$ dimensions. However, the latter contributes only 
perturbatively \mike.

The situation becomes slightly more complicated when one arrives at 
the step where a particular mode along the $n$-axis, say 
$\pm\omega_m ~(n=\pm m)$, is contained in $\Delta k$ and 
is to be eliminated. In this case, the RG
equation becomes
\eqn\mtwg{\eqalign{U_{\beta,k-\Delta k}(\Phi)-U_{\beta,k}(\Phi) &=
{1\over 2\beta}\sum_{n=-\infty}^{\infty}\int_{\sp}\biggl\{
\Theta\bigl(|\sp|
-k+\Delta k\bigr)\Theta\bigl(|n| < [{\bar\beta\over
{2\pi}}]\bigr)+\delta_{n,\pm m}\biggr\} \cr
&\qquad\qquad
\times\Theta\bigl(k-|\sp|\bigr)
~{\rm ln}\Bigl[\omega_n^2+p^2+U''_{\beta,k}(\Phi)\Bigr]\cr
&
={S_dk^{d-1}\over 2\beta}\bigl(\Delta k\bigr)\sum_{|n| \le
[{\bar\beta\over{2\pi}}]-1}
~{\rm ln}\Bigl[\omega_n^2+p^2+U''_{\beta,k}(\Phi)\Bigr] \cr
&
+{S_d\over\beta}\int_0^k dp~p^{d-1}~~{\rm ln}
\Bigl[\omega_m^2+p^2+U''_{\beta,k}(\Phi)\Bigr]\,,}}
where the last term represents the contribution from
$\pm \omega_m$. 
For $d=3$ we can evaluate the last term analytically
\eqn\thrp{\eqalign{
&{1\over 2\pi^2}\int_0^{k}dp~p^2{\rm ln}
\Bigl[\omega_{m}^2+p^2+U''_{\beta,k}(\Phi)\Bigr] \cr
&
={1\over 6\pi^2}\Biggl\{k^3{\rm ln}\Bigl[k^2+\omega_m^2
+U''_{\beta,k}\Bigr]+2\bigl[\omega_m^2+U''_{\beta,k}\bigr]^{3/2}
{\rm tan}^{-1}\Bigl({k\over{\sqrt{\omega_m^2+U''_{\beta,k}}}}\Bigr)\cr
&
+2k\bigl[\omega_m^2+U''_{\beta,k}(\Phi)\bigr]\Biggr\}\,.}}
The absence of the $\Delta k$ factor in the expression 
prevents us from generating a differential equation for $U_{\beta,k}(\Phi)$.
Nevertheless, Eq. \mtwg\ can be solved numerically. 

\medskip
\medskip
\item{(3)} $\rho_k^{(d)}(\sp,\omega_n)=\Theta(k-\sqrt{\omega_n^2
+|{\sp}|^2})$:
\medskip

The above smearing function represents another class of finite-temperature 
blocking, and has been considered in Refs.~\ref\roos\ and \ref\shafer.
In position space, it corresponds to
\eqn\smgr{ \rho_{k}^{(d)}(\sx,\tau_x)
={1\over\beta}\sum_{n=-\infty}^{\infty}\int_{\s p}
e^{-i(\omega_n\tau_x-\s p\cdot\s x)}~\Theta\bigl(k-\sqrt{\omega_n^2
+|{\sp}|^2}\bigr)\,.}
This particular choice of $\rho_k^{(d)}(\sp,\omega_n)$ leads to a mixing
between the components in the subspaces $S^1$ and $R^d$, and the Matsubara sum
over $n$ is now constrained by the value of $k$, similar to
Method 2 but in a more complicated manner. Figure 3 depicts how
modes contained in the RG shell are eliminated as 
$k$ is lowered.

\ifnum\dofig=1                                                       
\medskip                                                             
                                           
\centerline{\epsfbox{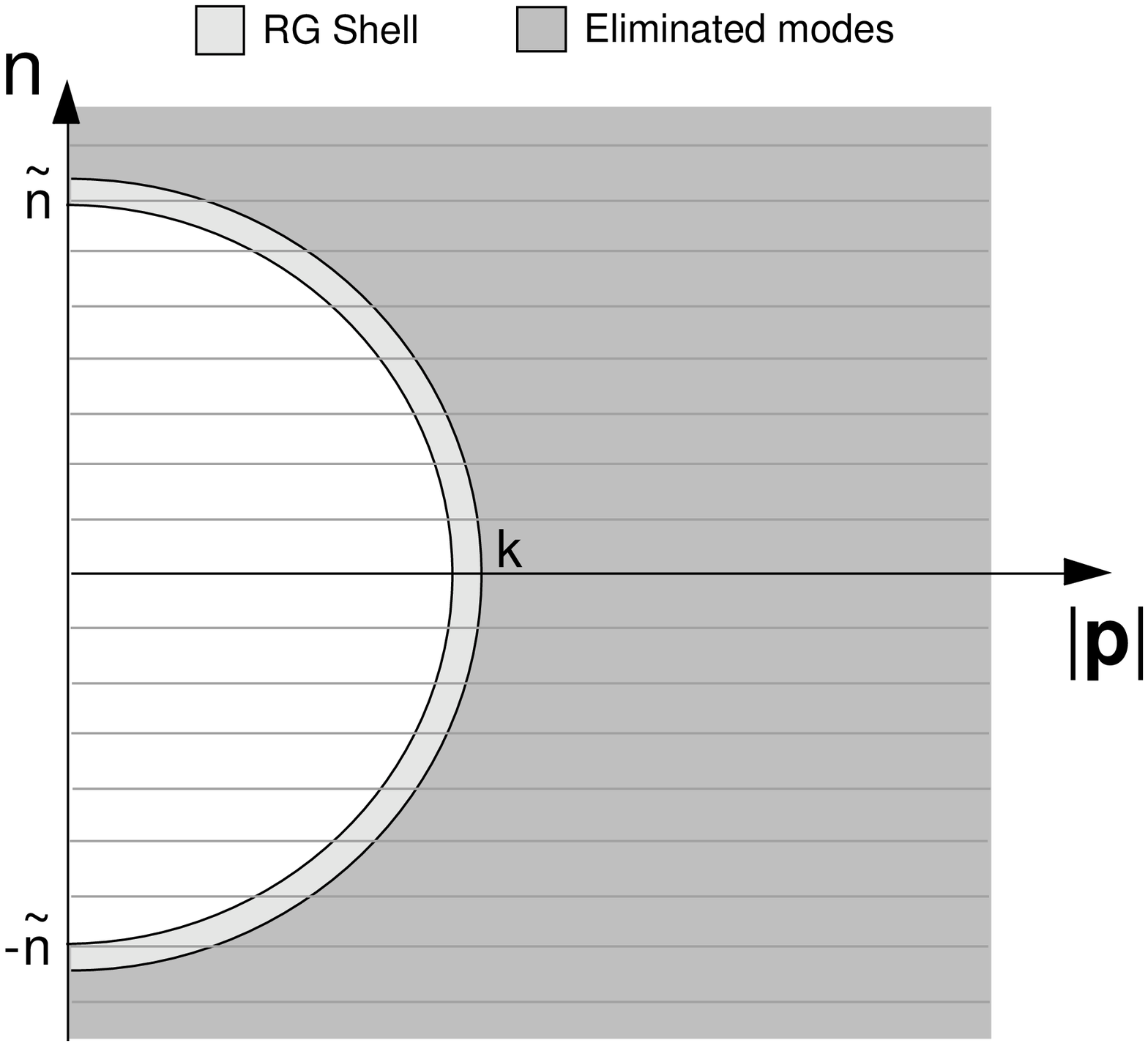}}                                      
\medskip                                                             
{\narrower                                                           
{\sevenrm                                                            
{\baselineskip=8pt  
\itemitem{Figure 3.}
Schematic diagram of blocking Method 3. 
\bigskip
}}}                                                                  
\fi 

If the RG shell does not contain a Matsubara mode along the $n$-axis,
a nonlinear differential flow equation can be obtained as
\eqn\srg{\eqalign{ k{{d U_{\beta,k}(\Phi)}\over{d k}} &=
-{k\over 2\beta}\sum_{n=-\infty}^{\infty}\int_{\sp} {\rm ln}\Bigl[\omega_n^2
+\sp^2+U''_{\beta,k}(\Phi)\Bigr]\delta\bigl(\sqrt{\omega_n^2
+|{\sp}|^2}-k\bigr) \cr
&
=-{S_dk^2\over 2\beta}\sum_n'\int_0^{\infty} dp~{p^{d-1}\over{
\sqrt{k^2-\omega_n^2}}}
~{\rm ln}\Bigl[\omega_n^2+p^2+U''_{\beta,k}(\Phi)\Bigr]\delta\bigl(
p-\sqrt{k^2-\omega_n^2}\bigr) \cr
&
=-{S_dk^2\over 2\beta}{\rm ln}\Bigl[k^2+U''_{\beta,k}(\Phi)
\Bigr]\sum_n'\bigl(k^2-\omega_n^2\bigr)^{{d-2}\over 2} \cr
&
=-{S_dk^d\over 2\beta}~g^{(d)}(\bar\beta)~{\rm ln}  
\Bigl[k^2+U''_{\beta,k}(\Phi)\Bigr]\,,}}
where 
\eqn\gfu{g^{(d)}(\bar\beta)=1+2\sum_{n=1}^{[\bar\beta/{2\pi}]}\Bigl[1-\bigl(
{{2\pi n}\over{\bar\beta}}\bigr)^2\Bigr]^{{d-2}\over 2}\,,}
is a piece-wise continuous function of $\bar\beta$, as depicted in 
Figure 4 for $d=3$. 

On the other hand, as in Method 2, an additional contribution representing
the $|m|$-th Matsubara mode along the $n$-axis appears during their
elimination, and the equation reads:
\eqn\mtwg{\eqalign{U_{\beta,k-\Delta k}(\Phi)& -U_{\beta,k}(\Phi) =
{1\over 2\beta}\sum_{n=-\infty}^{\infty}\int_{\sp}
\Theta\bigl(k-\sqrt{\omega_n^2+|\sp|^2}\bigr) \biggl\{
\Theta\bigl(\sqrt{\omega_n^2+|\sp|^2}-k+\Delta k\bigr)\cr
&\qquad\qquad
+\delta_{n,\pm m}
\Theta\bigl(\sqrt{\omega_n^2+|\sp|^2}-\omega_m\bigr)\biggr\} 
~{\rm ln}\Bigl[\omega_n^2+p^2+U''_{\beta,k}(\Phi)\Bigr]\cr
&
={S_dk^{d-1}\over 2\beta}\bigl(\Delta k\bigr)~{\tilde g}^{(d)}(\bar\beta)
~{\rm ln}\Bigl[k^2+U''_{\beta,k}(\Phi)\Bigr] \cr
&
+{S_d\over\beta}\int_0^{\sqrt{k^2-\omega_m^2}}dp~p^{d-1}~~{\rm ln}
\Bigl[\omega_m^2+p^2+U''_{\beta,k}(\Phi)\Bigr]\,,}}
where ${\tilde g}^{(d)}(\bar\beta)$ is the same as $g^{(d)}(\bar\beta)$
in \gfu\ except the $n=\pm [\bar\beta/{2\pi}]$ contribution is excluded in
the summation. Thus, we apply Eq.~\mtwg\ whenever a particular $|\omega_m|$
is contained in the RG shell. It could happen that the modes lie exactly 
at the boundary of the shell, i.e., $k-\Delta k=\omega_m$. In this case
one may easily verify that Eq. \mtwg\ reduces to Eq. \srg\ and a differential
flow equation is recovered. Although one could in principle 
fine tune $\Delta k$ with respect to $T$ in a way
such that the points along the $n$-axis always lie on the boundary of the
shells, we shall treat $\Delta k$ and $T$ as independent input parameters.

\ifnum\dofig=1                                                       
\medskip                                                             
                                           
\centerline{\epsfbox{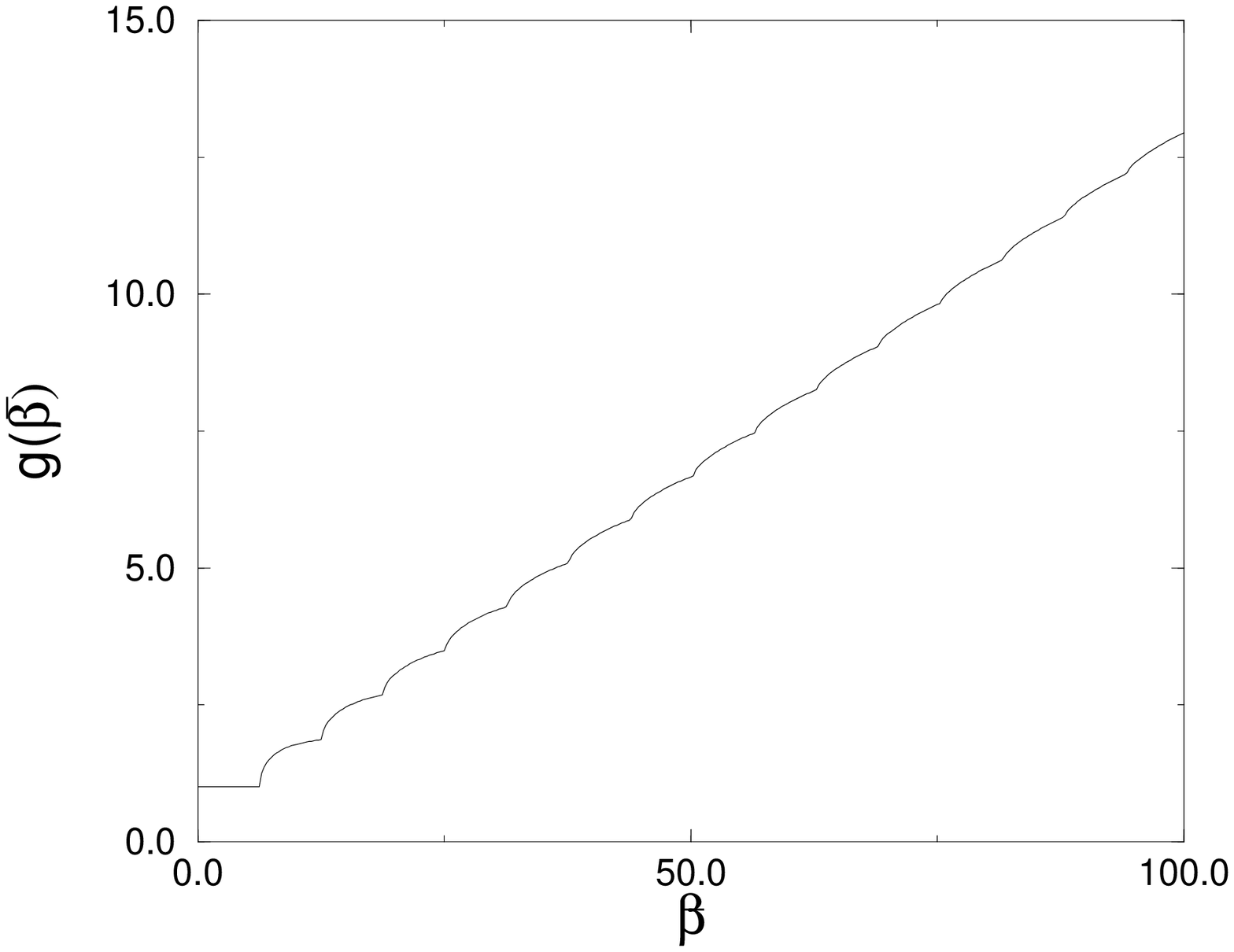}}                                      
\medskip                                                             
{\narrower                                                           
{\sevenrm                                                            
{\baselineskip=8pt  
\itemitem{Figure 4.}
Plot of $\scriptstyle g^{(3)}(\bar\beta)$ for Method 3.
\bigskip
}}}                                                                  
\fi 

The novel feature
of employing this type of blocking is that for both $\bar\beta \gg 1$ 
and $\bar\beta
\ll 1$, one will obtain the expected 
zero-temperature result as given in Eq. \ulfe,
namely, $D=d+1$ for $\bar\beta \gg 1$
and $D=d$ for $\bar\beta \ll 1$. 
This can be readily seen by noting that 
in the high-temperature limit $\bar\beta\to 0$ (decreasing $\beta$ while 
holding $k$ fixed), $g^{(d)}(\bar\beta)\to 1$ 
and Eq. \srg\ is reduced to Eq. \ulrg. On the other hand, in
the regime where $\bar\beta \gg 1$, the summation in Eq. \gfu\ can be replaced
by an integral: 
\eqn\gfus{\eqalign{ g^{(d)}(\bar\beta) & \approx 1+2\int_1^{\bar\beta
\over{2\pi}}
dn \Bigl[1-\bigl({{2\pi n}\over{\bar\beta}}\bigr)^2\Bigr]^{{d-2}\over 2} \cr
&
= 1+ {\sqrt\pi}~{\Gamma({d\over 2})\over{\Gamma({d+1\over 2})}}~
\Bigl({\bar\beta\over{2\pi}}\Bigr)
-2F\bigl({1\over 2},{1-{d\over 2}}, {3\over 2};
{4\pi^2\over{\bar\beta^2}}\bigr)
={S_{d+1}\over S_d}~\bar\beta+\cdots\,.}}
In this limit, $S^1\times R^d$ is effectively $R^{d+1}$ and 
the RG equation is reduced to that of \ulfe\ with $D=d+1$.

\medskip
\medskip
\item{(4)} $\rho_k^{(d)}(\sp,\omega_n)=\Theta(k-|\omega_n|-|{\sp}|)$:
\medskip

The above smearing function represents yet another method of blocking 
at finite temperature. The manner in which the modes are eliminated is
illustrated in Figure 5.

\ifnum\dofig=1                                                       
\medskip                                                             
                                           
\centerline{\epsfbox{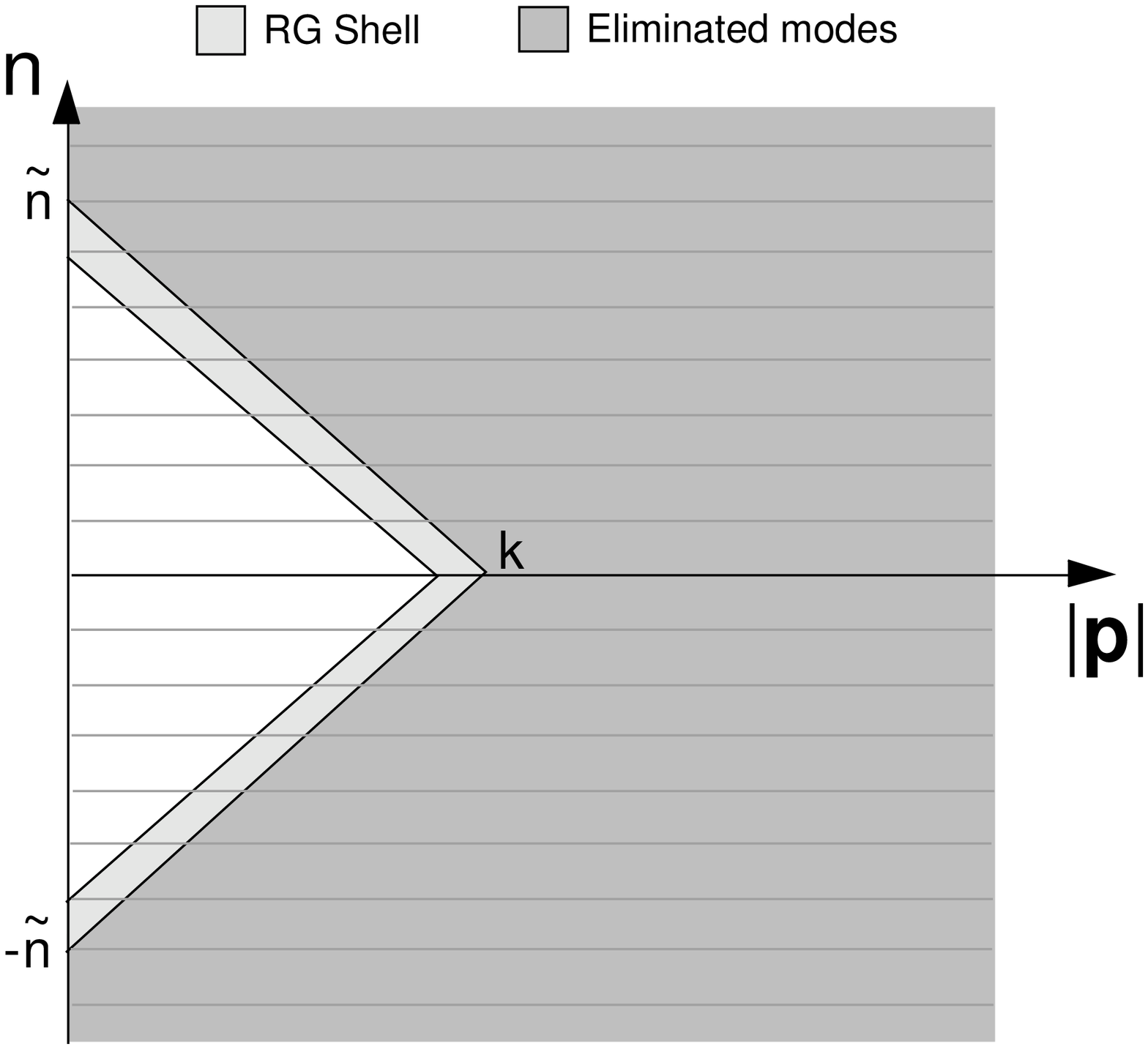}}                                      
\medskip                                                             
{\narrower                                                           
{\sevenrm                                                            
{\baselineskip=8pt                                                   
\itemitem{Figure 5.}
Schematic diagram of blocking Method 4.
\bigskip                                                             
}}}                                                                  
\fi 
In analogy to Method 3, the components $|\sp|$ and $\omega_n$ are 
also mixed due
to the constraint from the blocking function.
By comparing the results obtained from these
two schemes, one can gain insight into the effects of this mixing.
The RG equations can be written as 
\eqn\fsrg{\eqalign{ & k{{d U_{\beta,k}(\Phi)}\over{d k}} =
-{k\over 2\beta}\sum_{n=-\infty}^{\infty}\int_{\sp} {\rm ln}\Bigl[\omega_n^2
+\sp^2+U''_{\beta,k}(\Phi)\Bigr]\Biggl\{ \theta(n)\delta\bigl(\omega_n
+|{\sp}|-k\bigr) \cr
&
\qquad\qquad\qquad + \delta_{n,0}\delta\bigl(|{\sp}|-k\bigr)+ \theta(-n)
\delta\bigl(-\omega_n+|{\sp}|-k\bigr)\Biggr\} \cr
&
=-{S_dk^d\over 2\beta}\Biggl\{ 
2\sum_{n=1}^{[\bar\beta/{2\pi}]}
\Bigl(1-{{2\pi n}\over{\bar\beta}}\Bigr)^{d-1}
{\rm ln}\Bigl[\omega_n^2+(k-\omega_n)^2+U''_{\beta,k}(\Phi)
\Bigr]+{\rm ln}\bigl[k^2+U''_{\beta,k}(\Phi)\bigr]\Biggr\}\,,}}
and 
\eqn\mtd{\eqalign{ k\Biggl(&{{U_{\beta,k-\Delta k} -U_{\beta,k}}
\over{\Delta k}}\Biggr)={S_dk^d\over 2\beta}\Biggl\{ 
2\sum_{n=1}^{[\bar\beta/{2\pi}]-1}
\Bigl(1-{{2\pi n}\over{\bar\beta}}\Bigr)^{d-1}
{\rm ln}\Bigl[\omega_n^2+(k-\omega_n)^2+U''_{\beta,k}(\Phi)
\Bigr] \cr
&\qquad\qquad\quad
+{\rm ln}\bigl[k^2+U''_{\beta,k}(\Phi)\bigr]\Biggr\}
+{S_dk\over{\beta(\Delta k)}}\int_0^{k-\omega_m}dp~p^{d-1}~~{\rm ln}
\Bigl[\omega_m^2+p^2+U''_{\beta,k}(\Phi)\Bigr]\,,}}
for an RG shell with and without a Matsubara frequency mode 
$|\omega_m|$ along the
$n$-axis, respectively. Once more, Eqs \mtd\ and 
\fsrg\ coincide when the mode lies 
exactly along the boundary with $k-\Delta k=\omega_m$.

For $\bar\beta\to 0$, Eq. \ulrg\ is again easily recovered.
Similarly,
for $\bar\beta\to \infty$, $\omega_n$ can be regarded
as a continuous variable and
we replace the summation by an integral. Dropping the subscript 
$\beta$ in the potential, we obtain
\eqn\fsc{\eqalign{ k{{d U_k(\Phi)}\over{d k}} 
&=-{S_dk^d\over 2\beta}\Biggl\{ {\beta\over\pi}\int_{2\pi/{\beta}}^k
d\omega_n
\Bigl(1-{\omega_n\over k}\Bigr)^{d-1}
{\rm ln}\Bigl[\omega_n^2+\bigl(k-\omega_n\bigr)^2
+U''_k(\Phi)\Bigr] \cr
&\qquad\qquad\qquad
+{\rm ln}\Bigl[k^2+U''_k(\Phi)\Bigr]\Biggr\} \cr
&
= -{S_dk^{d+1}\over{2\pi}} \int^1_{2\pi/{\bar\beta}}dx\bigl(1
-x\bigr)^{d-1}
{\rm ln}\Bigl[x^2+\bigl(1-x\bigr)^2+{U''_k(\Phi)\over k^2}\Bigr]+\cdots \cr
&
\approx -{S_dk^{d+1}\over{2\pi}} {\rm ln}\Bigl[1+{U''_k(\Phi)\over k^2}
\Bigr]\int_0^1 dx\bigl(1-x\bigr)^{d-1} +\cdots\cr
&
=-{S_dk^{d+1}\over{2d\pi}} {\rm ln}\Bigl[1+{U''_k(\Phi)\over k^2}\Bigr]+
\cdots\,.}}
In going from the second to the third equation above, we have taken
the limit ${2\pi/{\bar\beta}}\to 0$. In addition, since the argument in 
the logarithm does not change appreciably with $x$ over the interval,  
we have dropped its $x$ dependence and moved the expression outside the
integral. While the expected logarithmic form of the RG flow is 
regained, the equation contains a factor $S_d/{d\pi}$ instead of the
expected $S_{d+1}$ for $D=d+1$ from Eq. \ulfe.
Such a constant overall mismatch, however, is harmless and 
can be eliminated by a simple rescaling:
$k\to c^{1/{(2-d)}}k$ and $U_k(\Phi)\to c^{2/{(2-d)}}U_k(\Phi)$
with $c=S_d/{S_{d+1}d\pi}$ for $d \ne 2$. 

Up to now we have discussed four types of smearing functions for setting
up the finite-temperature MRG. In the limit
$\bar\beta\to 0$, all four MRG equations reproduce the
expected dimensionally reduced result. The agreement can be easily understood
from the fact that at any arbitrary IR scale $k$, each of 
the four different RG shells
always contains the static $n=0$ mode which gives the dominant 
contribution for small $\bar\beta$, or the high-temperature limit.
The $n\ne 0$ modes are strongly suppressed and
can be treated perturbatively. In this regard, 
we expect to obtain qualitatively the same
physical information from any of the four RG equations in the
small $\bar\beta$ limit. 
We shall substantiate this claim later with numerical solutions.

On the other hand, the asymptotic forms of the four
MRG equations in the large $\bar\beta$ regime all differ from one another.
In Figure 6 the comparison between Methods 1 and 3 is depicted.
The difference may be attributed to
their treatment of couplings between $|\sp|$ and 
the $\omega_{n\ne 0}$ modes. While the $n=0$ sector alone has been shown
to be sufficient to attain proper scalings for $\bar\beta \ll 1$, to
reproduce  
the less trivial scalings in the $\bar\beta \gg 1$ regime will require a
subtle and an intricate account of the feedbacks between $|\sp|$ and these 
non-static modes. 
The analyses above show that only Method 3 is constructed in
such a way that it
leads to the desired low-temperature scalings for the running parameters,
while rescalings or adjustments of the UV cutoff are required for the
other methods.

\ifnum\dofig=1     
\medskip
\bigskip                                                  
\medskip                                                             
                                           
\centerline{\epsfbox{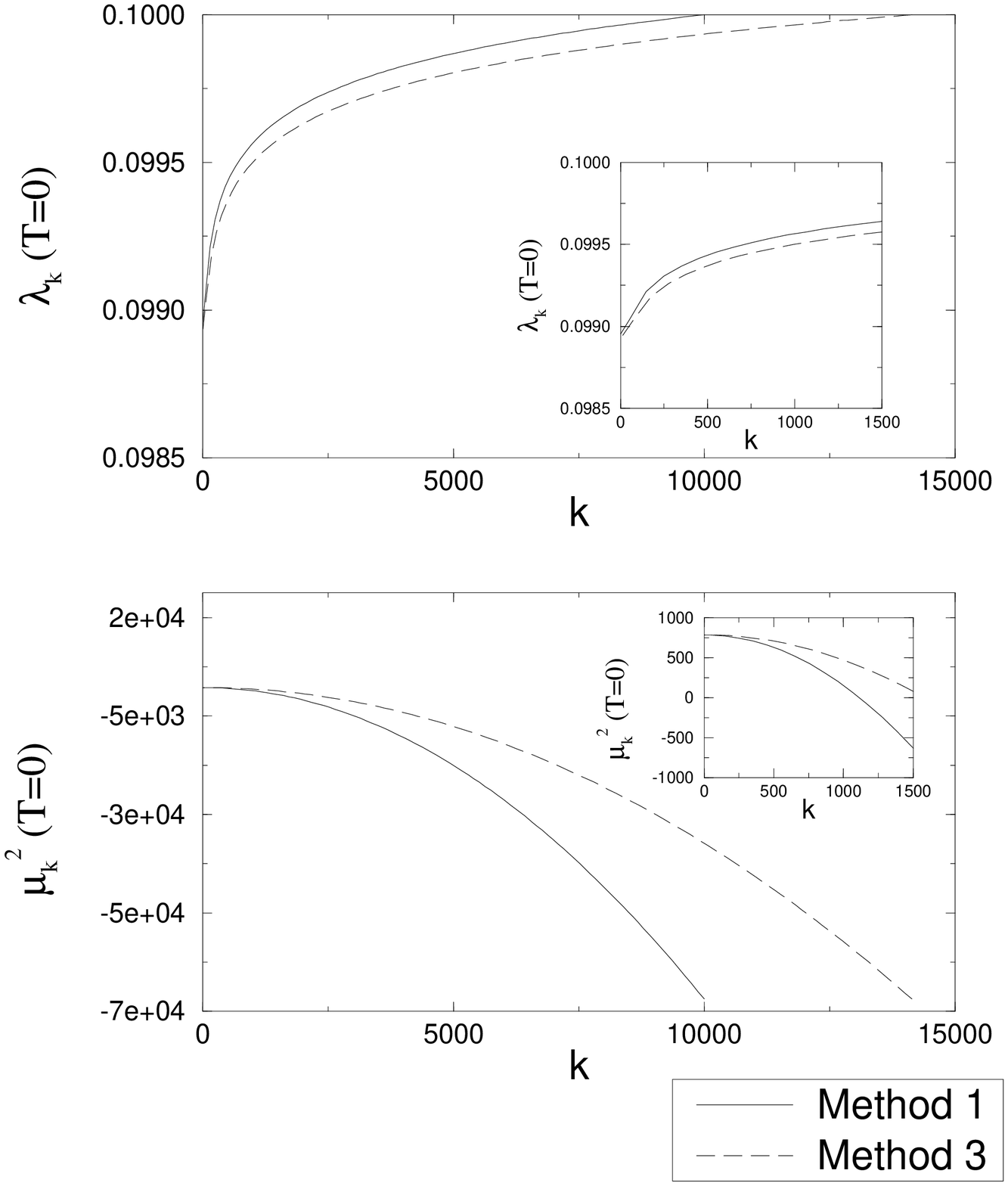}}                               
\medskip                                                             
{\narrower                                                           
{\sevenrm                                                            
{\baselineskip=8pt                                                   
\itemitem{Figure 6.}                                                 
Zero-temperature RG flow using Methods 1 and 3.  The figure shows that
as long as the correction to the UV cutoff is taken into account both methods
give the same renormalized couplings at $\scriptstyle{T=0}$.
\bigskip                                                             
}}}                                                                  
\fi                                                                  
\bigskip

In passing we comment that there exists a complementary TRG in which $\beta$ 
acts as
the running parameter. When $\beta$ is varied infinitesimally, the TRG 
leads to (see Appendix A) 
\eqn\ftrg{\eqalign{ \beta {\partial U_{\beta,k}}\over{\partial\beta}
&= -\int_{\s p}^{'}
\Biggl\{{1\over\beta}{\rm ln~sinh}\Bigl( {\beta\over 2} \sqrt{\sp^2+U''_{\beta,\sp}}  \Bigr) \cr
&
-{1\over 2}\Bigl[\sqrt{\sp^2+U''_{\beta,\sp}}
+\beta{\partial\over{\partial\beta}}\sqrt{\sp^2+U''_{\beta,\sp}}~\Bigr]
{\rm coth}~\Bigl({\beta \over 2} \sqrt{\sp^2+U''_{\beta,\sp}}
\Bigr)\Biggr\}\,,}}
with $k=\Lambda-\Delta k$ in the $|\sp|$ integration. This equation is
the finite-temperature analog of the renown Wegner-Houghton
RG \ref\wegner, and is consistent with the results obtained using the 
MRG schemes.
The general procedure of blocking transformation
is not unique, and one can certainly construct other types of smearing 
functions in addition to what have been considered. There is an 
important stipulation, however, for what is considered to be an effective
RG transformation -- that it makes the smallest possible change in going from
$k$ to $k-\Delta k$ for any arbitrary $k$. In other words, the most
successful scheme is the one whose RG shell contains the least number
of modes. Only by doing so is the environment not altered grossly. In
all four RG prescriptions, the fraction of the modes taken out from
each blocking step is $\kappa=\Delta k/k$. Since each loop 
is accompanied by a power of $\kappa$, by taking this parameter to be 
infinitesimally small all higher loop corrections can be suppressed 
\wegner, \ref\polchinski.
This justifies our MRG equations, which are based on a simple one-loop 
framework. On the other hand, the failure of perturbation theory and
the Matsubara shell assumption in the
high-temperature regime can be readily explained by noting that
$\kappa\approx 1$, which implies a dramatic change of the environment.

One could certainly consider a smearing function such as  
$\rho_{\tilde n}({\omega_n})=\Theta(\tilde n-|n|)$, where the RG
evolution is now governed by a
difference equation:
\eqn\nlio{U^{\{\tn-1\}}_{\beta}(\Phi)=U^{\{\tn\}}_{\beta}(\Phi)+{1\over\beta}
\int_{\s p}{\rm ln}\Bigl[\omega_{\tn}^2+\sp^2+U^{\{n\}''}_{\beta}(\Phi)
\Bigr]\,,}
or
\eqn\disrg{ U^{\{0\}}_{\beta}(\Phi)=U^{\{m\}}_{\beta}(\Phi)+{1\over \beta}
\sum_{n=1}^m
\int_{\s p}{\rm ln}\Bigl[\omega_{n}^2+\sp^2+U^{\{n\}''}_{\beta}(\Phi)\Bigr]\,.}
In this case eliminating $|\omega_m|$ implies taking away 
a fraction of modes, $\kappa'=2/(2m+1)$, from the theory. However,
$\kappa'$ is clearly not small at all for small $m$. 
Hence this blocking method
will break down in certain regimes.

\medskip
\bigskip
\centerline{\bf III. NUMERICAL RESULTS}
\medskip
\nobreak
\xdef\secsym{3.}\global\meqno = 1
\medskip
\nobreak

We now present the numerical results for the four MRG equations
discussed in Sec. II. By expanding
$U_{\beta,k}(\Phi)$ as a power series in $\Phi^2$
\eqn\expa{ U_{\beta,k}(\Phi)=\sum_{m=1}^{\infty}g^{(2m)}_{\beta,k}
{(\Phi^2)}^m\,,}
followed by a truncation 
at $\Phi^{14}$, we transform the
non-linear partial differential MRG equation
into a set of non-linear coupled ordinary differential equations for the 
couplings, 
$\{g^{(2m)}_{\beta,k}\}$. These 
equations are then integrated 
using a fourth-order, adaptive step-size, Runge-Kutta integrator.
We take $d=3$ for definiteness.

As pointed out in Section II, an entirely continuous RG flow is not obtained
when $|\sp|$ and $\omega_n$ are coupled.  In particular, the
continuous RG flow equations in the form of differential equations, 
e.g. (2.13), can be employed only when the RG shell to be eliminated
does not span a Matsubara mode.  However, when a Matsubara
mode, say $|\omega_m|$, is contained in the shell, a discrete, e.g. 
(2.15), version must
be used. Let the distance between $k$ and $\omega_m$ be $\delta$, where
$0 \le \delta \le \Delta k$.  
While it is possible to choose the integration bounds in a manner
such that the step which spans $|\omega_m|$ has $\delta=0$, i.e., 
$|\omega_m|$ lies exactly on the boundary of the RG shell, 
it would be interesting to 
examine the dependence of the flow pattern on $\delta$, 
over which the discrete flow equations are applied. 
For Methods 3 and 4, as $\delta \rightarrow 0$, the
discrete equations can be readily shown to reduce to their corresponding
differential equations. However, Method 2, has a ``pathology''
due to the horizontal sections of the RG shell and 
is therefore not possible to establish an equivalence
with its continuous counterpart.
In Fig. 7 we give a plot of the temperature dependence of the thermal mass 
parameter
as a function of $\delta$ using Method 3.  
The corrections vanish in the limit $\delta \rightarrow
0$. Since having a large $\delta$ will destroy the exactness of the RG flow
in the same way as having a large $\kappa$, one must always make
$\delta$ as small as possible if not zero.

\medskip
\ifnum\dofig=1                                                       
\medskip                                                             
                                           
\centerline{\epsfbox{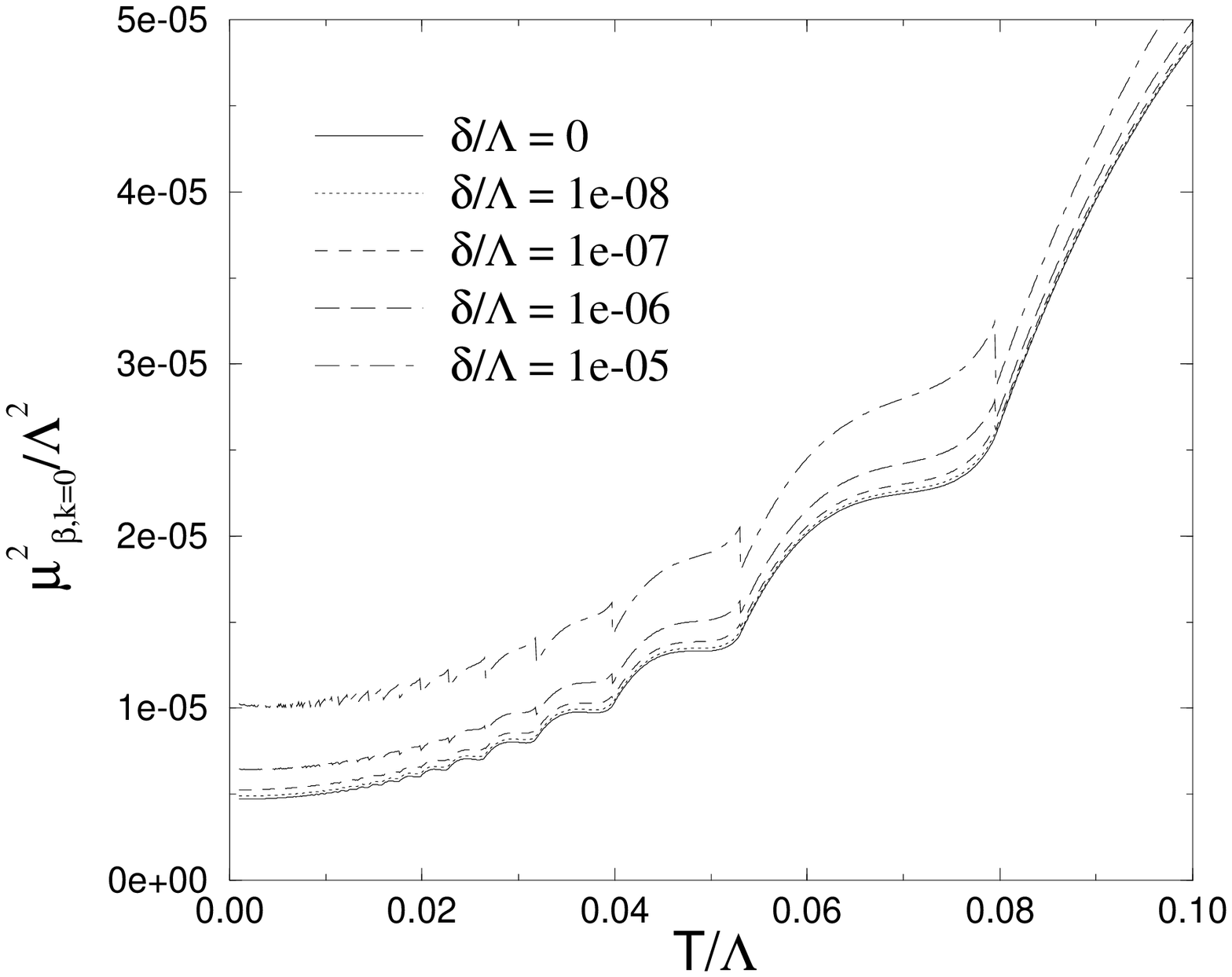}}                               
\medskip                                                             
{\narrower                                                           
{\sevenrm                                                            
{\baselineskip=8pt                                                   
\itemitem{Figure 7.}                                                 
Dependence of $\scriptstyle \mu^2_{\beta,k}$ on $\scriptstyle \delta.$ using Method 3.
\bigskip                                                             
}}}
\fi                                                                  
\bigskip

In Fig. 8 we plot the temperature dependence of the first six
coupling constants at $k=0$.
All four schemes show excellent agreement up to oscillations.
The oscillations obtained using Methods 2, 3, and 4 are due to the 
boundary conditions on the
RG flow.  For certain $T$, the constraint $\omega_{n_{max}} = \Lambda$ is 
satisfied exactly at $k=\Lambda$,  while for others it is not.  
Technically speaking
one should only compare data points which satisfy this constraint because
different boundary
conditions are mixed in the intermediate points and give rise to 
oscillations in the $g^{(2m)}_{\beta,k}$. The observed oscillations
are a numerical artifact of
the finiteness of the UV cutoff and signals the elimination of a particular
$|\omega_n|$. A smooth curve may be obtained for methods 3 and 4 if we 
choose our UV cutoff $\Lambda$ to be an integer multiple of $T$ and set 
$\Delta k=2\pi/{{\cal N}\beta}$ for some integer $\cal N$. 
Irrespective of this technicality, it is clear from Fig. 8
that all four methods give qualitatively the same results for the
temperature dependence of coupling constants.

\medskip
\ifnum\dofig=1                                                       
\medskip                                                             
                                           
\centerline{\epsfbox{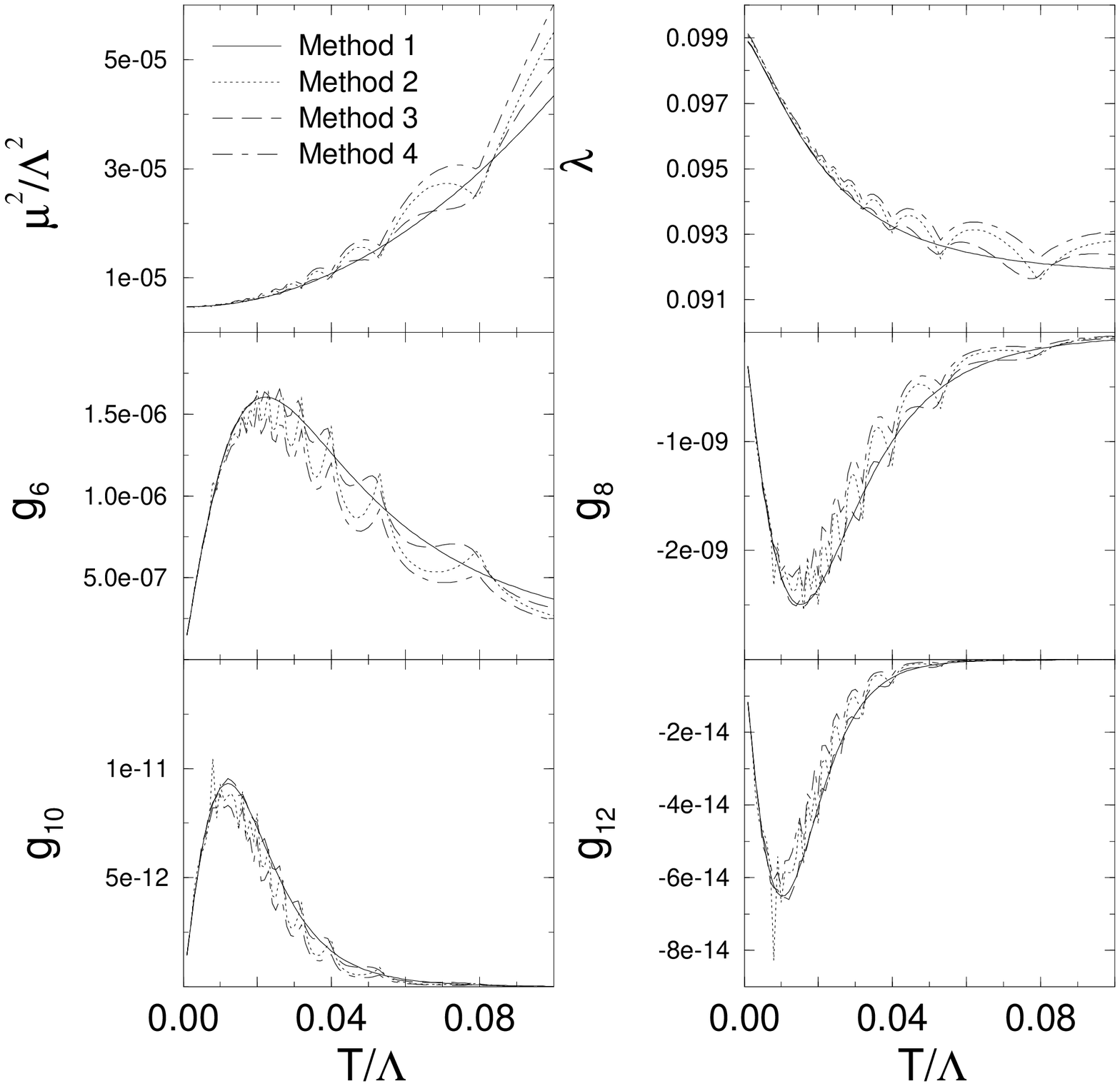}}                               
\medskip                                                             
{\narrower                                                           
{\sevenrm                                                            
{\baselineskip=8pt                                                   
\itemitem{Figure 8.}                                                 
Temperature dependence of $\scriptstyle g^{(2m)}$ in 
symmetric phase for $\scriptstyle 
m=1,\cdots 6$. The solid, dotted, dashed lines, and dot-dashed lines
are for Methods 1, 2, 3, and 4 respectively.
\bigskip                                                             
}}}                                                                  
\fi                                                                  
\bigskip
Let us now examine more closely the MRG flow of 
the relevant coupling constants, namely,
the mass parameter $\mu^2_{\beta,k}$ and the coupling constant 
$\lambda_{\beta,k}$.
At finite temperature, according to the prediction of the one-loop
perturbative approximation, the running mass parameter 
$\mu^2_{\beta}=\mu^2_{\beta,k=0}$
exhibits a quadratic dependence on $T$. Applying the four RG schemes
does not modify its strong $T^2$ dependence, i.e.,
incorporating the continuous feedback between the modes does not alter
the temperature dependence of the thermal mass in the high-$T$ limit.

The situation with the coupling constant $\lambda_{\beta}
=\lambda_{\beta,k=0}$, unfortunately, has not been so clear. Its
high-temperature behavior
has been a subject of recent debate \ms \roos. 
In this regime, a naive
one-loop calculation breaks down completely and
predicts that $\lambda_{\beta}$ decreases and turns negative eventually. 
Including the two-loop contribution does not help either.
Thus,
RG remains the only tool which could  
shed light on the true behavior of $\lambda_{\beta}$.
However, the results again differ depending on how RG is implemented and
at what scale the coupling constant is defined.  In Ref. \roos, for instance,
it was claimed that when using Method 3, $\lambda_{\beta}$ 
decreases at first, but
then increases logarithmically.  While the author attributed the discrepancy
between his result and that of Ref. \ms\ based on Method 1 to the 
functional form of the smearing functions, we find that  
this is not the case. The actual cause of the discrepancy lies in 
the scale at which the coupling constants are defined, namely, in the former,
$\lambda_{\beta, k=T}$ was considered, i.e.,
the coupling constant is defined at the scale $k=T$, whereas 
$\lambda_{\beta, k=0}$ was used in the latter. Had the 
normalization conditions agreed, the same result would have been obtained 
for all four RG schemes: $\lambda_{\beta,k=T}$ decreases, reaches
a minimum, and then increases logarithmically. However, 
defining the coupling constant at $k=0$ shows that
$\lambda_{\beta,k=0}$ decreases and
approximately approaches a constant for large $T$ for all four methods.

\medskip
\ifnum\dofig=1                                                       
\medskip                                                             
                                           
\centerline{\epsfbox{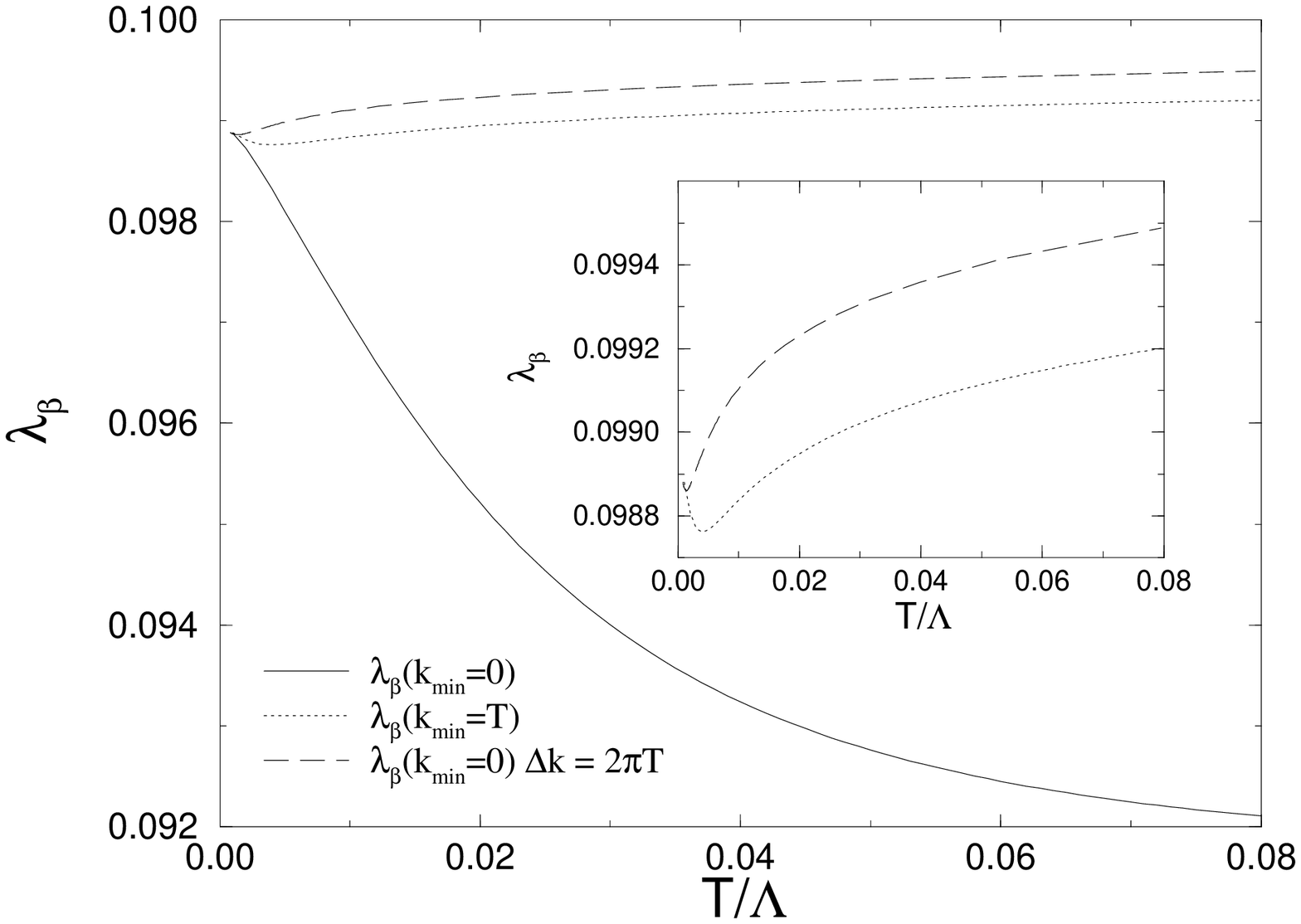}}                               
\medskip                                                             
{\narrower                                                           
{\sevenrm                                                            
{\baselineskip=8pt                                                   
\itemitem{Figure 9.}                                                 
Comparison of $\scriptstyle \lambda_{\beta,k}$ with different normalization
conditions.
\bigskip                                                             
}}}                                                                  
\fi                                                                  
\bigskip

Certain methods, however, will yield erroneous temperature dependence
of the coupling constants. An example is the Matsubara shell assumption
in which a shell of thickness $\Delta k=2\pi/{\beta}$ is integrated out
in each step of the RG evolution. This assumption gives $\kappa\sim 1$
in the high $T$ limit and the method breaks down completely (see Figure 9).
In addition, this method yields mean field (four-dimensional) critical
exponents instead of the Wilson-Fisher three-dimensional ones.  For
example, the exponent
$\beta$, defined by $\Phi_{\beta,k}\sim |T-T_c(k)|^{\beta}$ as $\beta\to
\beta_c(k)$, where $\Phi_{\beta,k}$ is the scale-dependent minimum of
the potential \mike, turns out to be $\beta = 1.01 \pm 0.01$ with the
Matsubara-shell assumption.

We comment that the $T$ dependence of the running parameters can also
be examined from the point of view of the RG trajectories and the 
fixed-point structure of the theory. The
presence of the infinite-temperature Gaussian fixed point 
implies that the RG trajectory must move
in such a way that the fixed-point 
$(\mu^{2*}_{\beta},\lambda_{\beta}^{*})=(\infty,0)$ is approached with
increasing $T$.
However, a careful analytical study reveals that the leading-order
increment of $\mu^2_{\beta}$ is $\lambda_{\beta}T^2$, which means that
the flow toward the
fixed point must be slowed down as $\lambda_{\beta}$ continues to
decrease according to the coupled relations (see Appendix B):
\eqn\gapmm{\eqalign{\mbb&=\mu_R^2+{\lb\over 24\beta^2}-{{\lb\bmb}\over 
8\pi\beta^2}-{\lb\mbb\over 16\pi^2}\Bigl[~{\rm ln}\Bigl({\bmb\over 4\pi}
\Bigr)+\gamma-1\Bigr] \cr
\lb&=\lambda_R-~{3\lb^2\over 16\pi\bmb}
-{3\lb^2\over 16\pi^2}\Bigl[~{\rm ln}\Bigl({\bmb\over 4\pi}\Bigr)
+\gamma\Bigr]\,,}}
with $\bar\mu_{\beta}=\beta\mu_{\beta}$.
The continuous feedback between $\mu^2_{\beta}$ and
$\lambda_{\beta}$ in turn yields a slower convergence
to their fixed point values. Since for a
realistic system the attainment of $\mu^2_{\beta}\to\infty$ is rather
unlikely, the corresponding $\lambda_{\beta}\to 0$ limit will not be
realized either. Therefore, it suffices to
approximate $\lambda_{\beta}$ by a constant. 

Further argument against the increase of $\lambda_{\beta}$ with 
$T$ can also be made
from the consideration of dimensional reduction, which takes
place only if $\bmb\to 0$ in the limit of vanishing $\beta$ \ref\ginsparg. 
As we have
seen before, the quantity $\bmb$ is of order $\lambda^{1/2}_{\beta}$.
Hence, if $\lambda_{\beta}$ increases with $T$, $\bar\mu_{\beta}$ will
also rise and the picture of high-temperature dimensional reduction will
be destroyed completely. 

\bigskip
\centerline{\sevenrm \vbox{\tabskip=0pt \offinterlineskip
\def\tablerule{\noalign{\hrule}}
\halign to375pt{\strut#&\vrule#\tabskip=1em plus2em&
  #\hfil& \vrule#& #\hfil& \vrule#&
  #\hfil& \vrule#&  #\hfil& \vrule# 	\tabskip=0pt\cr\tablerule
&& \omit\hidewidth $O(\Phi^{(2m)})$ 	\hidewidth
&& \omit\hidewidth $\gamma$		\hidewidth
&& \omit\hidewidth $\nu$ or $\zeta$ 	\hidewidth
&& \omit\hidewidth $\lim_{k \to 0} \lambda_{\beta_c,k} T_c/\mu_{\beta_c,k}$	\hidewidth
&\cr\tablerule
&&   2	  && 1.054 $\pm$ 0.005 	&& 0.527 $\pm$ 0.005 && 13.79 $\pm$ 0.05 &\cr\tablerule
&&   3	  && 1.171 $\pm$ 0.005	&& 0.585 $\pm$ 0.005 && 29.35 $\pm$ 0.05 &\cr\tablerule
&&   4	  && 1.345 $\pm$ 0.005	&& 0.672 $\pm$ 0.005 && 37.23 $\pm$ 0.05 &\cr\tablerule
&&   5	  && 1.504 $\pm$ 0.005	&& 0.744 $\pm$ 0.005 && 38.78 $\pm$ 0.05 &\cr\tablerule
&&   6	  && 1.352 $\pm$ 0.005	&& 0.675 $\pm$ 0.005 && 35.50 $\pm$ 0.05 &\cr\tablerule
&&   7	  && 1.312 $\pm$ 0.005	&& 0.655 $\pm$ 0.005 && 35.20 $\pm$ 0.05 &\cr\tablerule
}}}

\bigskip
{\narrower
{\sevenrm
{\baselineskip=7pt
\itemitem{Table 1.}
Critical exponents and the effective coupling constant 
as a function of the level of 
polynomial truncation.  Results for all four smearing functions are the 
same within the numerical accuracy.  Errors reported here 
are due to numerics only and not that 
resulting from the polynomial truncation
of the potential.
\bigskip
}}}

Other quantities can be measured to demonstrate the 
consistency of all four momentum blocking schemes.
For example, one may examine
the case $\mu^2_R< 0$ where the symmetry is spontaneously broken.
Although $\lambda_{\beta}$ vanishes at criticality, the effective
measure of the interaction is characterized, not by $\lambda_{\beta}$, but
by $\lambda_3 \equiv
\lim_{k\rightarrow 0} \lim_{T\rightarrow T_c}
\lambda_{\beta,k}T/\mu_{\beta,k}$ which
approaches a constant as $T\to T_c$. The effective coupling strength in
the vicinity of transition is obtained using the polynomial expansion
approximation up to $\Phi^{14}$ 
for MRG and the results are summarized in Table
1. We again find agreement in all four MRG equations within their numerical
accuracy, and the results are comparable to
those reported in Ref.~\ref\guida. 
In a similar manner, the critical
exponents $\gamma$ and $\nu$ were measured. From Table 1 we again see that
all methods are consistent with one another. However, we caution the 
readers that these results will be less accurate than that obtained
without polynomial truncation. This is because near criticality
field amplitude fluctuations are strong, but such truncation
applies only to the small fluctuation limit \mike, \ref\morris.

\goodbreak

\medskip
\bigskip
\centerline{\bf IV. SUMMARY AND DISCUSSIONS}
\medskip
\nobreak
\xdef\secsym{4.}\global\meqno = 1
\medskip
\nobreak

In this paper we have constructed finite-temperature RG equations
for a scalar field theory in $S^1\times R^d$ using a sharp momentum cutoff 
as the running parameter. 
We examined four different blocking schemes and illustrated how,
with an infinitesimally small thickness of the RG shell, $\Delta k$,
the feedback from the higher modes to the lower ones are systematically
incorporated as they are integrated out. The smallness of the parameter 
$\kappa={{\Delta k}/k}$, which represents the fraction of the modes that are
eliminated in each blocking step, is the most crucial ingredient for the
success of RG and constitutes
the basis for the ``exactness'' of 
Eqs.~\rgft, \urg, \srg\ and \fsrg,
since corrections from higher loops are strongly suppressed by additional 
powers of $\kappa$.

The four MRG equations differ essentially in their treatment of the 
feedbacks between
modes having the same $|\sp|$ but different $\omega_n$. 
In Method 1 the
feedbacks are ignored entirely, but the trade-off is that its 
unconstrained summation over
all $n$ leads to a continuous RG. On the other hand, Methods 2, 3 and 4 
which enforce certain couplings between $|\sp|$ and $\omega_n$ 
do not
lead to a differential function when there are non-vanishing contributions
from eliminating the modes along the $n$-axis. However,
for Methods 3 and 4, the differential forms are recovered 
by a judicious choice of
$\Delta k$ which puts all these points on the boundary of the RG shells. From 
the numerical results presented in
Sec.~III for the behavior of the $n$-point functions and the critical 
exponents, one finds remarkable agreement in all four methods, thus 
justifying the use of Method 1. We may therefore conclude that 
the couplings between different frequency modes with the same $|\sp|$ may be 
safely ignored, provided that $\kappa$ is chosen to be small. 
Had all the modes 
been strongly coupled, one would have to devise a laborious
scheme which eliminates one degree of freedom at a time.
Nevertheless, it remains
an interesting question to see why an ``independent-mode'' approximation for 
$n$ is sufficient for a small RG shell.

Knowing that physical
properties are independent of the methodology of blocking, we find Method 
1 to be most advantageous for numerical computation since it yields a 
continuous flow, with no oscillatory behavior in the high $T$ limit. 
On the other hand, to examine the scaling 
behavior of the running parameters in the low $T$ regime, Method 3 would be 
more convenient since it naturally leads back to the zero-temperature 
$(d+1)$-dimensional theory. 

Using the MRG (and TRG), we have examined the 
high-temperature behavior of coupling constant $\lambda_{\beta}$. 
Based on the 
fixed-point structure of the theory and the numerical results,
it is clear that $\lambda_{\beta,k=0}$ must decrease with $T$
but can 
be approximated by a positive constant in the limit $\bar {\mu_{\beta}} 
\ll 1$. The discrepancy in the literature concerning the behavior of
$\lambda_{\beta,k}$ can be reconciled with a precise renormalization
condition (viz, $k=0$ or $k=T$). 
We emphasize again that integrating out entire Matsubara shells 
at each RG step is ill-advised since it yields $\kappa\sim 1$ 
in the large $T$ limit
and the exactness of the RG flow is lost.

Various interesting directions can be 
pursued with our RG equations:
Numerous critical exponents have been measured by Method 1 with the 
inclusion
of wavefunction renormalization $Z_{\beta,k}(\Phi)$
in \mike. However, the accuracy in the polynomial
expansion can certainly be improved.
Derivations of the RG
equation for $Z_{\beta,k}(\Phi)$ using Methods 2, 3 and 4 can also
provide a consistency check with Method 1 \ref\derive.  
It would also be interesting to compare our predictions for  
the universal amplitude ratios with those obtained using the 
$\epsilon$ expansion and the lattice methods.
The one-component RG equation Eq. \rgft\ can readily be generalized to 
$N$ components:
\eqn\nrgft{\eqalign{ k{ {d U_{\beta, k}(\Phi)}\over dk} &=
-{S_dk^d\over 2\beta}\Biggl\{ \bigl(N-1\bigr)\biggl(
\beta\sqrt{k^2+U'_{\beta,k}(\Phi)}
+2~{\rm ln}\Bigl[1-e^{-\beta\sqrt{k^2+U'_{\beta,k}(\Phi)}}\Bigr]\biggr) \cr
&
+\beta\sqrt{k^2+U'_{\beta,k}(\Phi)
+U''_{\beta,k}(\Phi)\Phi^2}
+2~{\rm ln}\Bigl[1-e^{-\beta\sqrt{k^2+U'_{\beta,k}(\Phi)
+U''_{\beta,k}(\Phi)\Phi^2}}\Bigr]\Biggr\}\,,}}
where the prime notation now denotes the differentiation with respect to 
$\Phi^2/2=\Phi^a\Phi^a/2$ with $a=1,\cdots, N$. Critical exponents 
corresponding
to the XY $(N=2)$ model and Heisenberg $(N=3)$ model can now be extracted 
in a similar 
manner. In addition, the equation may also provide information on the 
interplay between the longitudinal and the transverse Goldstone modes. 
One may also explore different IR scaling regimes and search for the 
possible non-universal behavior which may arise due to the emergence of 
new relevant operators that are suppressed in the UV scaling regions 
\ref\polonyi. Nonperturbative modification of the scaling laws due to
the presence of a condesate can also be addressed.
Gauge theories, too, can also be further investigated \ref\gauge\ 
\ref\wkrg.
These issues will be presented in forthcoming publications.

\bigskip
\goodbreak
\bigskip
\centerline{\bf ACKNOWLEDGMENTS}
\medskip
\nobreak
S.-B. L. is grateful to J. Polonyi, V. Branchina, J. Alexandre and
A. Bodor for 
stimulating discussions. M.S. would like to thank D. Furnstahl and
E. Braaten for discussions as well.
This work is supported in part by funds provided by the National Science 
Council of Taiwan under contract \#NSC-87-2112-M-194-004, and by 
the National Science Foundation under Grants No. PHY-9511923 and PHY-9258270.

\medskip
\medskip
\centerline{\bf APPENDIX A. TEMPERATURE RENORMALIZATION GROUP}
\medskip
\nobreak
\xdef\secsym{{\rm A}.}\global\meqno = 1
\medskip
\nobreak
\medskip

In Sec II, the MRG prescription was explored 
from the point of view of momentum space
construction using an IR cutoff $k$ as the running parameter. The external
variable, temperature, has been taken as an input parameter. 
To examine the behavior of a physical system at finite temperature, there
exists an alternative temperature renormalization group (TRG),
in which $\beta$, the inverse of temperature, 
plays the role of the flow
parameter \ref\matsumoto. TRG provides a direct investigation of the response of a
system when $T$ is varied. Quantities at different $T$ and fixed $k$
can also be related to each other by this finite renormalization procedure. 

To construct the TRG flow equation, we again begin with the 
perturbative one-loop expression for the blocked potential given in 
Eq. \fubk. Differentiating the expression with 
respect to $\beta$ yields
\eqn\ftrgg{ {\beta}{{\partial\tilde U_{\beta,k}}\over{\partial\beta}}
=-\int_{\s p}^{'}\biggl\{{1\over\beta}{\rm ln~sinh}\Bigl({\beta\sqrt{
\sp^2+V''}
\over 2}\Bigr)-{\sqrt{\sp^2+V''}\over 2}~{\rm coth}
\Bigl({\beta\sqrt{\sp^2+V''}\over 2}\Bigr)\biggr\}\,.}
By varying $\beta$ infinitesimally from
$\beta\to\beta-\Delta\beta$, i.e., the system is heated up gradually until
the desired temperature scale is reached, we arrive at the following
improved TRG equation:
\eqn\ftrg{\eqalign{ \beta{{\partial U_{\beta,k}}\over{\partial\beta}}&=
-\int_{\s p}^{'}
\Biggl\{{1\over\beta}{\rm ln~sinh}\Bigl({{\beta\sqrt{\sp^2+U''_{\beta,\sp}} }
\over 2}\Bigr) \cr
&
-{1\over 2}\Bigl[\sqrt{\sp^2+U''_{\beta,\sp}}
+\beta{\partial\over{\partial\beta}}\sqrt{\sp^2+U''_{\beta,\sp}}~\Bigr]
{\rm coth}~\Bigl({{\beta \sqrt{\sp^2+U''_{\beta,\sp}}}
\over 2}\Bigr)\Biggr\}\,.}}
By assuming a small RG shell for the $|\sp|$ integration,
or equivalently, $k=\Lambda-\Delta k$, higher loop corrections again
are suppressed.

One can readily verify the equivalence between the TRG and the MRG
formalisms by observing that the same expression
\eqn\cdfr{\eqalign{k{\partial\over\partial k}\Bigl(\beta{{\partial 
U_{\beta,k}}
\over{\partial\beta}}\Bigr)&=S_dk^d
\Biggl\{{1\over\beta}{\rm ln~sinh}\Bigl({{\beta\sqrt{k^2+U''_{\beta,k}} }
\over 2}\Bigr) \cr
&
-{1\over 2}\Bigl[\sqrt{k^2+U''_{\beta,k}}
+\beta{\partial\over{\partial\beta}}\sqrt{k^2+U''_{\beta,k}}~\Bigr]
{\rm coth}~\Bigl({{\beta \sqrt{k^2+U''_{\beta,k}}}\over 2}\Bigr)\Biggr\}\,,}}
is obtained by differentiating \ftrg\ with respect to ${\rm ln}k$ or
\rgft\ with respect to ${\rm ln}\beta$. 
However, since Eq. \ftrg\ is an integro-differential 
equation, direct numerical solution is more cumbersome than
the MRG equations.
Notice that for large $\beta$, the
results derived from \ftrg\ and \ftrgg\ should not differ appreciably.
However, in the small $\beta$ regime, Eq.~\ftrgg, which is 
derived using
the independent-mode approximation, fails completely. 

The $\beta$ dependence of the running parameters in the coupling constant
space can be examined by differentiating \ftrg\ with respect to $\Phi$ and
setting $\Phi=0$. To $O(\lambda_{\beta,k})$, we have
\eqn\bmub{\eqalign{\beta{\partial\mbk\over{\partial\beta}}&=-{\beta\over 8}
\int_{\s p}^{'}{\rm csch}^2\Bigl({\sqrt{\sp^2+\mu^2_{\beta,\sp}}\over 2}\Bigr)
\Biggl\{\lambda_{\beta,\sp}
-{1\over {\sqrt{\sp^2+\mu^2_{\beta,\sp}}}}\Bigl({\partial\lambda_{
\beta,\sp}\over{\partial\beta}}\Bigr) {\rm sinh}\Bigl(
\sqrt{\sp^2+\mu^2_{\beta,\sp}}~\Bigr)\cr
&
+{\beta\lambda_{\beta,\sp}\over 2\bigl(\sp^2+\mu^2_{\beta,\sp}\bigr)}
\Bigl({\partial\mu^2_{\beta,\sp}\over{\partial\beta}}\Bigr)
\Bigl[1+{1\over{\beta\sqrt{\sp^2+\mu^2_{\beta,\sp}}
}}~{\rm sinh}\Bigl({\beta \sqrt{\sp^2+\mu^2_{\beta,\sp}}}~\Bigr)\Bigr]
\Biggr\}+{\rm C.T.},}}
\eqn\blamb{\eqalign{ \beta{\partial\lbk\over{\partial\beta}}&=-{3\beta^2
\over 16}\int_{\s p}^{'}{\lambda_{\beta,\sp}\over 
\sqrt{\sp^2+\mu^2_{\beta,\sp}}}
{\rm csch}^2\Bigl({{\beta \sqrt{\sp^2+\mu^2_{\beta,\sp}}}\over 2}\Bigr) 
\Biggl\{
{\lambda_{\beta,\sp}\over {\beta\sqrt{\sp^2+\mu^2_{\beta,\sp}}}} \cr
&
-\lambda_{\beta,\sp}
{\rm coth}\Bigl({{\beta\sqrt{\sp^2+\mu^2_{\beta,\sp}}}\over 2}\Bigr) 
+{2\over \sqrt{\sp^2+\mu^2_{\beta,\sp}}}\Bigl({\partial\lambda_{\beta,\sp}
\over{\partial\beta}}\Bigr)\cr
&
-{\beta\lambda_{\beta,\sp}\over\bigl(\sp^2+\mu^2_{\beta,\sp}\bigr)}
\Bigl({\partial\mu^2_{\beta,\sp}\over{\partial\beta}}\Bigr)
\biggl[{1\over{\beta \sqrt{\sp^2+\mu^2_{\beta,\sp}}}}
+{1\over 2}{\rm coth}\Bigl({{\beta \sqrt{\sp^2+\mu^2_{\beta,\sp}}}
\over 2}\Bigr)\biggr]\Biggr\}+{\rm C.T.},}}
where ${\rm C.T.}$ denotes the appropriate counterterms needed
to subtract off the divergences in the momentum integration.
Note that these counterterms 
are necessarily temperature-dependent \ref\oconnor.

Unfortunately, the coupled integro-differential equations are still too
difficult to be solved exactly; even analytical approximations are
restricted only to certain regimes. However, 
in the high-temperature regime where $\beta$ is small, we 
ignore the $\sp$ dependence in the mass and coupling constants and
obtain
\eqn\bmubb{\eqalign{ &\beta{{\partial{\mu^2_{\beta}}\over{\partial\beta}}}
\Biggl\{
1+{\lambda_{\beta}\beta\over 16}\int_{\s p}{1\over{\sp^2+\mu^2_{\beta}}} 
\Bigl[
1+{1\over {\beta\ssp }}{\rm sinh}\Bigl(\beta\ssp\Bigr)
\Bigr]{\rm csch}^2\Bigl({{\beta\ssp }\over 2}\Bigr)\Biggr\} \cr
&
=-{\lambda_{\beta}\beta\over 8}\int_{\sp}
{\rm csch}^2\Bigl({{\beta\ssp }\over 2}\Bigr)\Biggl\{\lambda_{\beta}
-{1\over\ssp}\Bigl({\partial\lambda_{\beta}\over{\partial\beta}}
\Bigr){\rm sinh}\Bigl(\beta\ssp\Bigr)\Biggr\}+{\rm C.T.},}}
and
\eqn\blambb{\eqalign{ \beta{\partial\lb\over{\partial\beta}} & \Biggl\{
1+{3\lambda_{\beta}\beta\over 8}\int_{\s p}{1\over{\sp^2+\mu^2_{\beta}}}
{\rm csch}^2\Bigl({{\beta\ssp }\over 2}\Bigr)\Biggr\} \cr
&
=-{3\lambda_{\beta}^2\beta^2\over 16}\int_{\sp}{1\over\ssp }
{\rm csch}^2\Bigl({{\beta\ssp }\over 2}\Bigr) 
\Biggl\{{1\over {\beta\ssp }}
-{\rm coth}\Bigl(\beta\ssp \Bigr) \cr
&
-{1\over{\sp^2+\mu^2_{\beta}}}\biggl[{1\over{\beta\ssp }}+{1\over 2}
~{\rm coth}\Bigl(\beta\ssp\Bigr)\biggr]\Biggr\}+{\rm C.T.}\,,}}
for vanishing $k$. The above two RG equations will
allow us to examine the fixed-point structure of the theory, as we
shall see in Appendix B.

\goodbreak

\medskip
\medskip
\centerline{\bf APPENDIX B. RENORMALIZATION GROUP AND THE FIXED POINTS}
\medskip
\nobreak
\xdef\secsym{{\rm B}.}\global\meqno = 1
\medskip
\nobreak
\medskip

The complicated RG flow can be greatly simplified in the
neighborhood of fixed points where a linearized RG prescription becomes
possible and only the relevant operators need to be retained.
The irrelevant operators can
be safely ignored. The
classification of relevancy, however, depends on the fixed point;
what is relevant around a particular fixed point may turn irrelevant as it
flows into another one. Thus, the linearized prescription may be used
only when the system is sufficiently close to a fixed point.
To describe the flow further away from the fixed point,
a global nonlinear RG analysis is generally required. 
Once all the fixed points are located, the general RG flow of the theory
can be mapped out. Below we examine the fixed-point structure of both
MRG and TRG.

\medskip
\medskip
\centerline{\bf 1. MRG Approach}
\medskip
\medskip

For MRG, we shall work with Method 1 since its RG evolution is continuous.
It is convenient to employ the dimensionless quantities: 
\eqn\dimmm{\eqalign{&\bar U_{\beta,k}(\bar\Phi)=\beta k^{-d}U_{\beta,k}
(\Phi),\quad\quad \bar\Phi=\beta^{1/2}k^{-(d-2)/2}\Phi,\cr
&\bar U^{(m)}_{\beta,k}(\bar\Phi)={{\partial^m\bar U_{\beta,k}(\bar\Phi)}\over
{\partial\bar\Phi^m}}=\beta^{1-m/2}k^{-d+m(d-2)/2}U^{(m)}_{\beta,k}(\Phi),\cr
&\bar\mu^2_{\beta,k}=\bar U^{(2)}_{\beta,k}(0)=k^{-2}\mu^2_{\beta,k},~~
\quad \bar\lambda_{\beta,k}=\bar U^{(4)}_{\beta,k}(0)=\beta^{-1} k^{d-4}
\lambda_{\beta,k}\,,}}
where, for the purpose of the present paper,
we set the anomalous dimension $\eta$ to be zero.
Eq. \rgft\ can now be written as  
\eqn\drgft{\biggl[k\partial_k-{1\over 2}\bigl(d-2\bigr)\bar\Phi
\partial_{\bar\Phi}+d\biggr]\bar U_{\beta,k}
=-{S_d\over 2}\Biggl\{\bar\beta\sqrt{1+\bar U''_{\beta,k}}+2{\rm ln}
\bigl[1-e^{-\bar\beta\sqrt{1+\bar U''_{\beta,k}} }\bigr]\Biggr\}\,,} 
which implies the following fixed-point equation:
\eqn\fprg{ -{1\over 2}\bigl(d-2\bigr)\bar\Phi \bar U_{\beta,k}^{*'}
+d\bar U^{*}_{\beta,k}
=-{S_d\over 2}\Biggl\{\bar\beta\sqrt{1+\bar U^{*''}_{\beta,k}}+2{\rm ln}
\bigl[1-e^{-\bar\beta\sqrt{1+\bar U^{*''}_{\beta,k}} }\bigr]\Biggr\}\,,} 
with $\bar U_{\beta,k}^{*}(\bar\Phi)$ being the fixed-point potential.
However, by making a polynomial expansion of the blocked
potential $U_{\beta,k}(\bar\Phi)$ and keeping track of only the evolution
of the leading-order ${\bar\mu}^2_{\beta,k}$ and ${\bar\lambda}_{\beta,k}$, 
it is
possible to identify the fixed points on the $({\bar\mu}^2_{\beta,k},
{\bar\lambda}_{\beta,k})$ plane.  Consistency with reflection symmetry around
the origin demands that only even-power terms are present and we have
\eqn\uexpa{\bar U_{\beta,k}(\bar
\Phi)=\sum_{m=1}^{\infty}{
{\bar g}^{(2m)}_{\beta,k}\over{(2m)!}}\bar\Phi^{2m},\qquad\qquad 
{\bar g}^{(2m)}_{\beta,k}={\bar U}^{(2m)}_{\beta,k}(0)\,,}
with ${\bar g}^{(2)}_{\beta,k}=\bar\mu^2_{\beta,k}$ and 
and ${\bar g}^{(4)}_{\beta,k}=\bar\lambda_{\beta,k}$.

In the high-temperature limit where $\bar\beta \ll 1$, the manifold changes
from $S^1\times R^d$ to $R^{d}$, exhibiting a critical behavior governed 
by the $d$-dimensional fixed point. Thus, one expects
\eqn\resca{\bar\beta\to 0:\quad \cases{\eqalign{& S^1\times R^d\longrightarrow 
R^d \cr
&\bar\Phi\longrightarrow \bar\Phi_d~~~~~~~~~~~ \bigl(\Phi\longrightarrow
\beta^{-1/2}\Phi_d\bigr) \cr
&\bar U_{\beta,k}\longrightarrow \bar U_{k,d}~~~~~~\bigl(U_{\beta,k}
\longrightarrow \beta^{-1} U_{k,d}\bigr)\cr
&\bar g^{(m)}_{\beta,k}\longrightarrow\bar g^{(m)}_{k,d}~~~~~~\bigl( 
g^{(m)}_{\beta,k}\longrightarrow\beta^{m/2-1}g^{(m)}_{k,d}\bigr)\,, \cr}}}
and
\eqn\diyd{\biggl[k\partial_k-{1\over 2}\bigl(d-2\bigr)
\bar\Phi_d\partial_{\bar\Phi_d}+d\biggr]\bar U_{k,d}(\bar\Phi_d)
=-{S_d\over 2}{\rm ln}\Biggl[{{1+\partial^2_{\bar\Phi_d}
\bar U_{k,d}(\bar\Phi_d)}\over{1+\partial^2_{\bar\Phi_d}\bar U_{k,d}
(0)}}\Biggr]\,.}
Concentrating only on the running of $\bar\mu^2_{\beta,k}$ and 
$\bar\lambda_{\beta,k}$, the evolution equations 
can be approximated as
\eqn\muhg{\eqalign{ k{\partial{\bar\mu}^2_{k,d}\over {\partial k}}
&=-2{\bar\mu}^2_{k,d}-{{
\bar\lambda_{k,d} S_d}\over2(1+\bar\mu^2_{k,d})} \cr
k{\partial{\bar\lambda}_{k,d}\over{\partial k}} 
&=-(4-d)\bar\lambda_{k,d}
+{3\bar\lambda_{k,d}^2S_d\over2(1+\bar\mu^2_{k,d})^2}\,.}}
%
%
%
If we further define
\eqn\mbafd{ \bar r={{\bar\mu^2_{k,d}}\over{1
+\bar\mu^2_{k,d}}}, \qquad\qquad
\bar\ell={\bar\lambda_{k,d}\over(1+\bar\mu^2_{k,d})^2}\,,}
Eq.~\muhg\ then simplifies to
\eqn\muggf{\eqalign{k{\partial{\bar r}\over{\partial k}}
&=-(1-\bar r)\bigl(2\bar 
r+{S_d\over 2}\bar\ell\bigr) \cr
k{\partial{\bar\ell}\over{\partial k}} &=-\bar\ell\Bigl(\epsilon-4\bar r
-{5S_d\over 2}\bar\ell\Bigr)\,,}}
where $\epsilon=4-d$. Fixed points can now be located.  There are two
trivial fixed points of Gaussian nature located at $(\bar
r^{*},\bar\ell^{*})=(0,0)$ and $(1,0)$. The former is the usual trivial
``finite'' Gaussian fixed point at which all running parameters vanish. On
the other hand, the latter describes an infinite (temperature) Gaussian
fixed point which can only be reached when $T\to\infty$.  In addition, for
$D=d < 4$ or $\epsilon > 0$, we have the IR Wilson-Fisher (WF)
fixed point $(\bar r^{*},\bar\ell^{*}) =(-\epsilon/6,2\epsilon/3S_d)$
\ref\zinn, which coincides with the usual trivial
Gaussian fixed point for $\epsilon=0$. Thus, we have 
$\bar\ell^{*}={16\pi^2\epsilon/3}$ for $d=4$.
While Eq.~\muggf\ is already diagonal
around the infinite Gaussian fixed point, it is possible to introduce the
new variables \ref\nicoll
\eqn\newv{\bar x={3S_d\over 2}\bar\ell,\qquad\qquad
\bar y=\bar r+{S_d\over 2(2-\epsilon)}\bar\ell\,,}
and transform \muggf\ into
\eqn\finga{\eqalign{k{\partial{\bar x}\over{\partial k}}
&=-\bar x\Bigl[\epsilon
\bigl(1-\bar x\bigr)-4\bar y\Bigr]\cr
k{\partial{\bar y}\over{\partial k}}&=2\bar y\bigl(\bar y
-1-{{\epsilon\bar x}\over 6}\bigr)+O(\epsilon^2\bar x^2)\,.}}
The new expression is now diagonalized around the finite Gaussian fixed
point $(0,0)$, with the infinite Gaussian and the WF fixed points located,
respectively, at $(0,1)$ and $(1,0)$ on the $(\bar x,\bar y)$ plane. 

On the other hand, in the low $T$ limit, one expects
\eqn\rescal{\bar\beta\to\infty:\quad \cases{\eqalign{& S^1\times R^d
\longrightarrow R^{d+1} \cr
&\bar\Phi\longrightarrow \bar\beta^{1/2}\bar\Phi_{d+1}~~~~~~~~~~~~~~ 
\bigl(\Phi\longrightarrow\Phi_{d+1}\bigr) \cr
&\bar U_{\beta,k}\longrightarrow \bar\beta^{-1}\bar U_{k,d+1}~~~~~~~~~
\bigl(U_{\beta,k}\longrightarrow  U_{k,d+1}\bigr)\cr
&\bar g^{(m)}_{\beta,k}\longrightarrow\bar\beta^{1-m/2}\bar g^{(m)}_{k,d+1}
~~~~~~\bigl(g^{(m)}_{\beta,k}\longrightarrow g^{(m)}_{k,d+1}\bigr)\,, \cr}}}
and 
\eqn\cfdsa{ \biggl[k\partial_k-{1\over 2}\bigl(d-2\bigr)\bar\Phi
\partial_{\bar\Phi}+d\biggr]\bar U_{k,d+1}(\bar\Phi)
=-{S_d\bar\beta\over 2}\sqrt{1+\partial^2_{\bar\Phi}\bar U_{k,d+1}
(\bar\Phi)}\,,} 
which yields 
\eqn\ghtr{\eqalign{k{\partial{\bar\mu}^2_{k,d+1}\over{\partial k}}
&=-2\bar\mu^2_{k,d+1}
-{\bar\lambda_{k,d+1}S_d\over 4(1+\bar\mu^2_{k,d+1})^{1/2}} \cr
k{\partial{\bar\lambda}_{k,d+1}\over{\partial k}}&=-(3-d)\bar\lambda_{k,d+1}
+{3\bar\lambda^2_{k,d+1}S_d\over 8(1+\bar\mu^2_{k,d+1})^{3/2}}\,.}}
Here we see that the only fixed point is the trivial finite Gaussian
$(\bar\mu^{2*}_{k,d+1},\bar\lambda^{*}_{k,d+1})=(0,0)$, for $d \ge 3$.
While the finite Gaussian fixed point can be accessed in both low and high
$T$ regimes, the WF and the infinite Gaussian fixed points are reachable
only in the large $T$ limit. The rationale for the inability to detect
these two fixed points at low $T$ is that they are characteristic of $R^d$
with $d < 4$, and yet the effective manifold is $R^{d+1}$ for small $T$. 

\medskip
\medskip
\centerline{\bf 2. TRG Approach}
\medskip
\medskip

The fixed-point
structure associated with the flow of TRG can be examined in a similar 
manner. We focus again only on the evolution of the leading-order
running coupling constants $\mu^2_{\beta}$ and $\lambda_{\beta}$ at $k=0$.
To simplify the task of solving the coupled integro-differential equations
Eqs. \bmubb\ and \blambb, we choose the manifold to be $S^1\times R^3$ for
definiteness. 

In the limit $\bar\beta\mu_{\beta} \ll 1$, the RG equations can be
approximated as
\eqn\cdgh{\eqalign{  \beta{\partial\mu^2_{\beta}\over{\partial\beta}}&={
\mu_{\beta}^2\over 4\pi^2}\Biggl\{ \lambda_{\beta}
\bmb{\partial F_1(\bmb)\over{\partial\bmb}}+\beta{\partial\lambda_{\beta}
\over\partial\beta}F_1(\bmb)\Biggr\} \cr
&
=-{1\over 12\beta^2}\Biggl\{\lambda_{\beta}\Bigl(1-
{\bmb\over 2\pi}+{3\bmb^2\over 4\pi^2}\Bigr)
-\beta{{\partial\lambda_{\beta}}\over{\partial\beta}}\Bigl(
{1\over 2}-{3\bmb\over 2\pi}\Bigr)+\cdots\Biggr\}\,,}}
and
\eqn\geuh{\beta{\partial\lambda_{\beta}\over\partial\beta}=-{3
\lambda_{\beta}^2\beta\over 16\pi^2\mu_{\beta}}
\Bigl(\beta{\partial\mu_{\beta}^2\over\partial\beta}+2\mu_{\beta}^2\Bigr)
{{\partial F_2(\bmb)}\over{\partial\bmb}} 
=-{3\lambda_{\beta}^2\over 32\pi^2\mu^2_{\beta}}
\Bigl(\beta{\partial\mu_{\beta}^2\over\partial\beta}+2\mu_{\beta}^2\Bigr)
\Bigl(1-{\pi\over\bmb}\Bigr)+\cdots\,.}
where 
\eqn\cfgh{\eqalign{&\int_{\sp}{\rm csch}^2\Bigl({{\beta\ssp }
\over 2}\Bigr)
=-{2\mu_{\beta}^3\over\pi^2}{{\partial F_1(\bmb)}\over{\partial\bmb}}, \cr
&\int_{\sp}{1\over\ssp }{\rm coth}\Bigl({{\beta\ssp }
\over 2}\Bigr)={\mu_{\beta}^2\over\pi^2}F_1(\bmb)+\cdots\,,}}
and \ref\hweldon\
\eqn\cfer{\eqalign{F_1(\bmb)&=\int_0^{\infty}{dy~y^2\over\sqrt{y^2+1}}{1\over
{e^{\bmb\sqrt{y^2+1}}-1}}={2\pi^2\over\bmb^2}\Bigl[{1\over 12}-{\bmb\over 4\pi}
-{\bmb^2\over 8\pi^2}~{\rm ln}\bmb+\cdots\Bigr],\cr
{{\partial F_1(\bmb)}\over{\partial\bmb}}&=-\int_0^{\infty}dy{y^2e^{\bmb
\sqrt{y^2+1}}\over\bigl(e^{\bmb\sqrt{y^2+1}}-1\bigr)^2}=-{\pi^2\over 3\bmb^3}
\Bigl[1-{\bmb\over 2\pi}
+{3\bmb^2\over 4\pi^2}+\cdots\Bigr], \cr
F_2(\bmb)&=\int_0^{\infty}{dy\over\sqrt{y^2+1}}{1\over
{e^{\bmb\sqrt{y^2+1}}-1}}={1\over 2}~{\rm ln}\bmb+{\pi\over 2\bmb}+\cdots,\cr
{{\partial F_2(\bmb)}\over{\partial\bmb}}&=-\int_0^{\infty}dy{e^{\bmb
\sqrt{y^2+1}}\over\bigl(e^{\bmb\sqrt{y^2+1}}-1\bigr)^2}
={1\over 2\bmb}\bigl(1-{\pi\over \bmb}\bigr)
+\cdots\,.}}
Note that $\beta$ is now the running parameter of the RG.
One may actually derive the above evolution 
equations by differentiating the coupled self-consistent equations
\eqn\gapmm{\eqalign{\mbb&=\mu_R^2+{\lb\over 24\beta^2}-{{\lb\bmb}\over 
8\pi\beta^2}-{\lb\mbb\over 16\pi^2}\Bigl[~{\rm ln}\Bigl({\bmb\over 4\pi}
\Bigr)+\gamma-1\Bigr] \cr
\lb&=\lambda_R-~{3\lb^2\over 16\pi\bmb}
-{3\lb^2\over 16\pi^2}\Bigl[~{\rm ln}\Bigl({\bmb\over 4\pi}\Bigr)
+\gamma\Bigr]\,,}}
which are valid for a small but positive $\bar\mu_{\beta}$ \mike. 

Upon substituting the familiar leading-order result 
\eqn\cdff{ \beta{\partial\mu^2_{\beta}\over{\partial\beta}}=-{
\lambda_{\beta}\over 12\beta^2}+\cdots,\qquad{\rm or}\qquad
\mu^2_{\beta}=\mu_R^2+{\lb\over 24\beta^2}+\cdots\,,}
into Eq. \geuh, the fixed points may now be located by solving 
\eqn\fpoi{0=\beta{\partial\lb\over\partial\beta}=-{3\lb^2\over
16\pi^2}\Bigl(1-{\lb\over{24\bmb^2}}\Bigr)
\Bigl(1-{\pi\over\bmb}\Bigr)\,.}
Since the equation is derived for $\bar\mu_{\beta} \ll 1$, we require
$\bmb< \pi$. Two fixed points may now be located. The
first one is the usual trivial finite Gaussian fixed point
$(\mu_{\beta},\lb)=(0,0)$.  As for the second one, we see that for
$\lb/{24\bmb^2}\to 1$, or equivalently $\beta\to 0$, the theory approaches
the infinite Gaussian fixed point $(\mu_{\beta},\lb)=(\infty,0)$.  Its
Gaussian nature is readily seen by noting that $\bmb\sim\lb^{1/2}$, and
thus, when $\mb\to\infty$ and $\beta\to 0$ such that $\bmb\to 0$, $\lb$
must be vanishingly small. 

In general one may regard phase transitions as 
a ``high-temperature'' phenomenon
since $\mu_{\beta}\to 0$ in the vicinity of criticality.
Although we are able to detect two Gaussian fixed points, the important WF
fixed point which characterizes the critical behavior of the system
seems to be missing. 
A crucial observation here is that in the
large $T$ regime one expects the effective degrees of freedom to crossover
from four-dimensional to three-dimensional, and hence $\lambda_{\beta}$ is
not an accurate measure of the effective interaction strength of the
theory. Instead, the interaction is accounted for by
$\lambda_{\beta,3}=\lambda_{\beta}T$, or
$\bar\lambda_{\beta,3}=\lambda_{\beta}/{\bmb}$ in the dimensionless form.
Thus, in terms of $\bar\mu_{\beta}$ and $\bar\lambda_{\beta,3}$, we have
\eqn\flob{\eqalign{\beta{\partial\bar\mu_{\beta}\over{\partial\beta}}&
=\bar\mu_{\beta}-{\bar\lambda_{\beta,3}\over 24} \cr
\beta{\partial\bar\lambda_{\beta,3}\over{\partial\beta}}&
=-\bar\lambda_{\beta,3}\Bigl(1-{\bar\lambda_{\beta,3}\over{24\bar\mu_{\beta}}}
\Bigr)\Bigl[1+{3\bar\lambda_{\beta,3}\bar\mu_{\beta}\over{16\pi^2}}
\bigl(1-{\pi\over\bar\mu_{\beta}}\bigr)\Bigr]\,.}}
The much richer structure associated with the flow of
$\bar\lambda_{\beta,3}$ now reveals the existence of the additional
nontrivial fixed point.  Besides the usual finite $(0,0)$ and infinite
Gaussian $(1,0)$, we have, keeping only the leading-order contribution in
the bracket,
\eqn\fval{(\bar\mu_{\beta}^{*},\bar\lambda_{\beta,3}^{*})=
({2\pi\over 9},~{16\pi\over 3}),}
%
The numerical value for $\bar\lambda_{\beta,3}^{*}$ can be compared with
that tabulated in \guida. Therefore, contrary to the well known notion
of triviality for $T=0$ at dimensionality $D=4$, the theory in the high $T$
limit is seen to crossover to an interacting theory characterized by a
nontrivial fixed point of $D=3$. 

We comment that in the TRG formalism, by first taking the limit $k\to 0$
in Eqs. \bmub\ and \blamb, we are left with one arbitrary parameter
$\beta$, the circumference of the submanifold $S^1$, with all other
dimensions being infinite. Since any TRG transformation now formally
corresponds to changing the value of $\beta$, it is
natural to expect the extrema $\beta=0$ and $\beta=\infty$ to be
fixed points that characterize, respectively, the three-dimensional and
four-dimensional limits of the theory.  While $D=4$ for $\bmb
\gg 1$, it becomes three in the limit $\bmb \ll 1$. Therefore, the
dimensional crossover scale in this TRG formalism can be given
qualitatively as $\bmb\sim 1$. This implies that when $\bmb$ falls below
unity new effective three-dimensional degrees of freedom need to be
employed in order to give a correct account of the theory. However, had
the new degrees of freedom associated with $D=3$ not been used,
the WF fixed point would be missed. In fact, the success of a given RG
scheme in characterizing the crossover phenomenon is attributed to its
capability of tracking the effective degrees of freedom. We also note that
the method of resumming daisy and superdaisy diagrams proposed by Dolan
and Jackiw \ref\jackiw\ is similar to TRG; however, the important
distinction is that in the former approach, the coupling constant has
always been fixed at its zero-temperature renormalized value and
consequently, the prescription only gives the two Gaussian fixed points
and yields the mean-field exponents. Thus, we see that only when the
running of the effective coupling constant $\bar\lambda_{\beta,k}$ is
taken into account can one access the nontrivial WF fixed point. 

\medskip
\bigskip

\medskip
\medskip
\centerline{\bf REFERENCES}
\medskip
\medskip
\nobreak

\item{\lp} S.-B. Liao, J. Polonyi and D. P. Xu, {\it Phys. Rev.}
{\bf D51} (1995) 748.
\medskip
\item{\others} see, for example, D. Dalvit and F. Mazzitelli, {\it Phys. Rev.}
{\bf D54} (1996) 6338;
\medskip 
M. D'Attanasio and M. Pietroni,  {\it Nucl. Phys.} {\bf B472} (1996) 711; 
\medskip
K. Ogure and J. Sato, hep-ph/9801439 and hep-ph/9802418;
\medskip
N. Tetradis and C. Wetterich, {\it Nucl. Phys.} {\bf B422} (1994) 541. 
\medskip
\item{\wilson} K. Wilson, {\it Phys. Rev.} {\bf B4} (1971) 3174;
K. Wilson and J. Kogut, {\it Phys. Rep.} {\bf 12C} (1975) 75.
\medskip
\item{\ms} S.-B. Liao and M. Strickland, {\it Phys. Rev.}
{\bf D52} (1995) 3653.
\medskip
\item{\mike} S.-B. Liao and M. Strickland, {\it Nucl. Phys.} {\bf B497}
(1997) 611.
\medskip
\item{\sb} S.-B. Liao and J. Polonyi, {\it Ann. Phys.} {\bf 222} (1993) 122
and {\it Phys. Rev.} {\bf D51} (1995) 4474.
\medskip
\item{\patkos} A. Patkos, P. Petreczky and J. Polonyi, {\it Ann. Phys.}
{\bf 247} (1996) 78.
\medskip
\item{\roos} T. Roos, {\it Phys. Rev.} 
{\bf D54} (1996) 2944.
\medskip
\item{\shafer} J. D. Shafer and J. R. Shepard, {\it Phys. Rev.} 
{\bf D55} (1997) 4990.
\medskip
\item{\wegner} F.J. Wegner and A. Houghton, {\it Phys. Rev} 
{\bf A8} (1972) 401.
\medskip
\item{\polchinski} see, for example, J. Polchinski, {\it Nucl. Phys.} 
{\bf B231} (1984) 269;
\medskip
A. Hasenfratz and P. Hasenfratz, {\it ibid.} {\bf B270} (1986) 687.
\medskip
\item{\ginsparg} P. Ginsparg, {\it Nucl. Phys.} {\bf B170} (1980) 388.
\medskip
\item{\guida} R. Guida and J. Zinn-Justin, {\it Nucl. Phys.} {\bf B489} 
(1997) 626.
\medskip
\item{\morris} 
T. R. Morris, {\it Int. J. Mod. Phys.} {\bf A9} (1994) 2411; 
{\it Phys.Lett.} {\bf B334} (1994) 355 and {\bf B329} (1994) 241.
\medskip
\item{\derive} see, for example, C. M. Fraser, {\it Z. Phys.} 
{\bf C28} (1985) 101; 
\medskip
A. Bonanno and D. Zappala, hep-th/9712038.
\medskip
\item{\polonyi} J. Alexandre, V. Branchina and J. Polonyi, hep-th/9709060 
and  hep-th/9712147.
\medskip
\item{\gauge} S.-B. Liao, {\it Phys. Rev.} {\bf D53} (1996) 2020, and
{\it ibid.} {\bf D56} (1997) 5008.
\medskip
\item{\wkrg} see, for example, M. Bonini, M. D'Attanasio and G. Marchesini,
{\it Nucl. Phys.} {\bf B437} (1995) 163 and {\bf B409} (1993) 441;
\medskip
M. Reuter and C. Wetterich, {\it ibid.} {\bf B417} (1994) 181;
\medskip
J. Comellas and A. Travesset, {\it ibid.} {\bf B498} (1997) 539.
\medskip
\item{\matsumoto} H. Matsumoto, Y. Nakano and H. Umezawa, 
{\it Phys. Rev.} {\bf D29} (1984) 1116.
\medskip
\item{\oconnor} D. O'Connor and C. R. Stephens, {\it Int. J.
Mod. Phys.} {\bf A9} (1994) 2805.
\medskip
\item{\zinn} J. Zinn-Justin, {\it Quantum Field Theory and Critical
Phenomena}, (Clarendon, Oxford, 1989).
\medskip
\item{\nicoll} J. F. Nicoll, T. S. Chang and H. E. Stanley, {\it Phys. Rev.
Lett.} {\bf 32} (1974) 1446, {\bf 33} (1974) 540; {\it Phys. Rev.}
{\bf A13} (1976) 1251.
\medskip
\item{\hweldon} H. A. Weldon, {\it J. Math. Phys.} {\bf 23} (1982) 1852.
\medskip 
\item{\jackiw} L. Dolan and R. Jackiw, {\it Phys. Rev.} 
{\bf D9} (1974) 3320.
\medskip
\vfill
\eject
%
%
%
%
%
\medskip
\centerline{\bf FIGURE CAPTIONS}
\bigskip

\itemitem{Figure 1.}
Schematic diagram of blocking Method 1.
\medskip
\itemitem{Figure 2.}
Schematic diagram of blocking Method 2. The thickness of the RG shell
in the $n$-axis is zero for $ \Delta k \to 0$ 
(exaggerated here), but a Matsubara mode is eliminated after $\cal N$ 
continuous iterations where 
$[{\cal N}\beta\Delta k/{2\pi}]=1$.
\medskip
\itemitem{Figure 3.}
Schematic diagram of blocking Method 3. 
\medskip
\itemitem{Figure 4.}
Plot of $g^{(3)}(\bar\beta)$ for Method 3.
\medskip
\itemitem{Figure 5.}
Schematic diagram of blocking Method 4.
\medskip
\itemitem{Figure 6.}
Zero-temperature RG flow using Methods 1 and 3.  The figure shows that
as long as the correction to the UV cutoff is taken into account both methods
give the same renormalized couplings at ${T=0}$.
\medskip
\itemitem{Figure 7.}
Dependence of $\scriptstyle \mu^2_{\beta,k}$ on $\scriptstyle \delta.$ using Method 3.
\medskip
\itemitem{Figure 8.}
Temperature dependence of $ g^{(2m)}$ in 
symmetric phase for $ 
m=1,\cdots 6$. The solid, dotted, dashed lines, and dot-dashed lines
are for Methods 1, 2, 3, and 4 respectively.
\medskip
\itemitem{Figure 9.}
Comparison of $\lambda_{\beta,k}$ with different normalization
conditions.
\medskip
\bigskip
\centerline{\bf TABLE}
\bigskip

\itemitem{Table 1.}
Critical exponents and the effective coupling constant 
as a function of the level of 
polynomial truncation.  Results for all four smearing functions are the 
same within the numerical accuracy.  Errors reported here 
are due to numerics only and not that 
resulting from the polynomial truncation
of the potential.

\vfill
\eject
\end